\documentclass[aps,pre,twocolumn,groupedaddress,showkeys]{revtex4}
\usepackage{graphics,color}

\usepackage[]{graphicx}
\DeclareGraphicsExtensions{.pdf}

\newcommand {\e}[1]{\mathrm{~#1}}       
\newcommand {\E}[1]{\cdot 10^{#1}}		
\newcommand {\vek}[1]{\mathbf{#1}}

\begin{document}

\title{Searching for collective behavior in a network of real neurons}

\author{Ga\v{s}per Tka\v{c}ik$^{a}$, Olivier Marre$^{b,d}$, Dario Amodei$^{d,e}$, Elad Schneidman$^{c}$,  William Bialek,$^{e,f}$ and Michael J. Berry II$^{d}$}
\affiliation{$^a$Institute of Science and Technology Austria, A-3400 Klosterneuburg, Austria,\\
$^b$Institut de la Vision, INSERM U968, UPMC, CNRS U7210, CHNO Quinze-Vingts, F-75012 Paris, France,\\
$^c$Department of Neurobiology, Weizmann Institute of Science, 76100 Rehovot, Israel,\\
$^d$Department of Molecular Biology, 
$^e$Joseph Henry Laboratories of Physics, and
$^f$Lewis--Sigler Institute for Integrative Genomics,
Princeton University, Princeton, New Jersey 08544 USA}

\date{\today}

\begin{abstract}
Maximum entropy models are the least structured probability distributions that exactly reproduce a chosen set of  statistics measured in an interacting network. Here we use this principle to construct probabilistic models which describe the correlated spiking activity of populations of up to 120 neurons in the salamander retina as it responds to natural movies. Already in  groups as small as 10 neurons,  interactions between spikes can no longer be regarded as small perturbations in an otherwise independent system; for 40 or more neurons pairwise interactions need to be supplemented by a global interaction that controls the distribution of synchrony in the population. Here we show that such ``K-pairwise'' models---being systematic extensions of the previously used pairwise Ising models---provide an excellent account of the data. 
We explore the properties of the neural vocabulary by: 1) estimating its entropy, which constrains the population's capacity to represent visual information; 2) classifying activity patterns into a small set of metastable collective modes; 3) showing that the neural codeword ensembles are extremely inhomogenous; 4) demonstrating that the state of individual neurons is highly predictable from the rest of the population, allowing the capacity for error correction.

\end{abstract}

\maketitle

\section{Introduction}

Physicists have long hoped that the functional behavior of large, highly interconnected  neural networks could be described by statistical mechanics \cite{hopfield_82,amit_89, hertz_91}.   The goal of this effort has been not to simulate the details of particular networks, but to understand how interesting functions can emerge, collectively, from large populations of neurons.  The hope, inspired by our quantitative understanding of collective behavior in systems near thermal equilibrium, is that such emergent phenomena will have some degree of universality, and hence that one can make progress without knowing all of the microscopic details of each system.  A classic example of work in this spirit is the Hopfield model of associative or  content--addressable memory \cite{hopfield_82}, which is able to recover the correct memory from any of its subparts of sufficient size. Because the computational substrate of neural states in these models were binary ``spins,'' and the memories were realized as locally stable states of the network dynamics, methods of statistical physics could be brought to bear on theoretically challenging issues such as the storage capacity of the network or its reliability in the presence of noise \cite{amit_89, hertz_91}.   On the other hand, precisely because of these abstractions, it has not always been clear how to bring the predictions of the models into contact with experiment.

Recently it has been suggested that the analogy between statistical physics models and neural networks can be turned into a  precise mapping, and connected to experimental data,  using the maximum entropy framework \cite{schneidman+al_06}.  In a sense, the maximum entropy approach is the opposite of what we usually do in making models or theories.  The conventional approach is to hypothesize some dynamics for the network we are studying, and then calculate the consequences of these assumptions; inevitably, the assumptions we make will be wrong in detail.  In the maximum entropy method, however, we are trying to strip away all our assumptions, and find models of the system that have {\em as little structure as possible} while still reproducing some set of experimental observations.  

The starting point of the maximum entropy method for neural networks is that the network could, if we don't know anything about its function, wander at random among all possible states.  We then take measured, average properties of the network activity as constraints, and each constraint defines some minimal level of structure.  Thus, in a completely random system neurons would generate action potentials (spikes) or remain silent with equal probability, but once we measure the mean spike rate for each neuron we know that there must be some departure from such complete randomness.  Similarly, absent any  data beyond the mean spike rates, the maximum entropy model of the network is one in which each neuron spikes independently of all the others, but once we measure the correlations in spiking between pairs of neurons, an additional layer of structure is required to account for these data.  The central idea of the maximum entropy method is that, for each experimental observation that we want to reproduce, we add only the minimum amount of structure required.  

An important feature of the maximum entropy approach is that the mathematical form of a maximum entropy model is exactly equivalent to a problem in statistical mechanics.  That is, the maximum entropy construction defines an ``effective energy'' for every possible state of the network, and the probability that the system will be found in a particular state is given by the Boltzmann distribution in this energy landscape.  Further, the energy function is built out of terms that are related to the experimental observables that we are trying to reproduce.  Thus, for example, if we try to reproduce the correlations among spiking in pairs of neurons, the energy function will have terms describing effective interactions among pairs of neurons.  As explained in more detail below, these connections are not analogies or metaphors, but precise mathematical equivalencies.

Minimally structured models are attractive, both because of the connection to statistical mechanics and because they represent the absence of modeling assumptions.  Of course, these features  do not guarantee that such models will provide an accurate description of a real system.  Interest in maximum entropy approaches to networks of real neurons was triggered by the observation that, for groups of $N=10-15$ ganglion cells in the vertebrate retina, maximum entropy models based on the mean spike probabilities of individual neurons and correlations between pairs of cells indeed generate successful predictions for the probabilities of all the combinatorial patterns of spiking and silence  in the network as it responds to naturalistic sensory inputs \cite{schneidman+al_06}.   In particular, the maximum entropy approach made clear that genuinely collective  behavior in the network can be consistent with relatively weak correlations among pairs of neurons, so long as these correlations are widespread, shared among most pairs of cells in the system.  This approach has now been used to analyze the activity in a variety of neural systems \cite{shlens+al_06,tkacik+al_06,yu+al_08,tang+al_08,tkacik+al_09,shlens+al_09,ohiorhenuan+al_10,ganmor+al_11,vasquez+al_12,sdme}, the statistics of natural visual scenes \cite{tkacik+al_10b,stephens+al_13,saremi+al_13}, the structure and activity of biochemical and genetic networks \cite{lezon+al_06,tkacik_07},  the statistics of amino acid substitutions in protein families  \cite{bialek+ranganathan_07,seno+al_08,weigt+al_09,halabi+al_09,mora+al_10,marks+al_11,sulkowska+al_12},  the rules of spelling in English words \cite{stephens+bialek_10}, and the directional ordering in flocks of birds \cite{bialek+al_12}.

One of the lessons of statistical mechanics is that systems with many degrees of freedom can behave in qualitatively different ways from systems with just a few degrees of freedom.  If we can study only a handful of neurons (e.g., $N\sim 10$ as in Ref \cite{schneidman+al_06}), we can try to extrapolate based on the hypothesis that the group of neurons that we analyze is typical of a larger population.   These extrapolations can be made more convincing by looking at a population of $N=40$ neurons, and within such larger groups one can also try to test more explicitly whether the hypothesis of homogeneity or typicality is reliable \cite{tkacik+al_06,tkacik+al_09}.  All these analyses suggest that, in the salamander  retina, the roughly 200 interconnected neurons that represent a small patch of the visual world should exhibit dramatically collective behavior.  In particular, the states of these large networks should cluster around local minima of the energy landscape, much as for the attractors in the Hopfield model of associative memory \cite{hopfield_82}.  Further, this collective behavior means that responses will be substantially redundant, with the behavior of one neuron largely predictable from the state of other neurons in the network; stated more positively, this collective response allows for pattern completion and error correction.  Finally, the collective behavior suggested by these extrapolations is a very special one, in which the probability of particular network states, or equivalently the degree to which we should be surprised by the occurrence of any particular state, has an anomalously large dynamic range \cite{mora+bialek_11}.  If correct, these predictions would have a substantial impact on how we think about coding in the retina, and about neural network function more generally.  Correspondingly, there is some controversy about all these issues \cite{nirenberg,roudi1,roudi2,roudi3}.

\begin{figure}[bt]
\includegraphics[width=\linewidth]{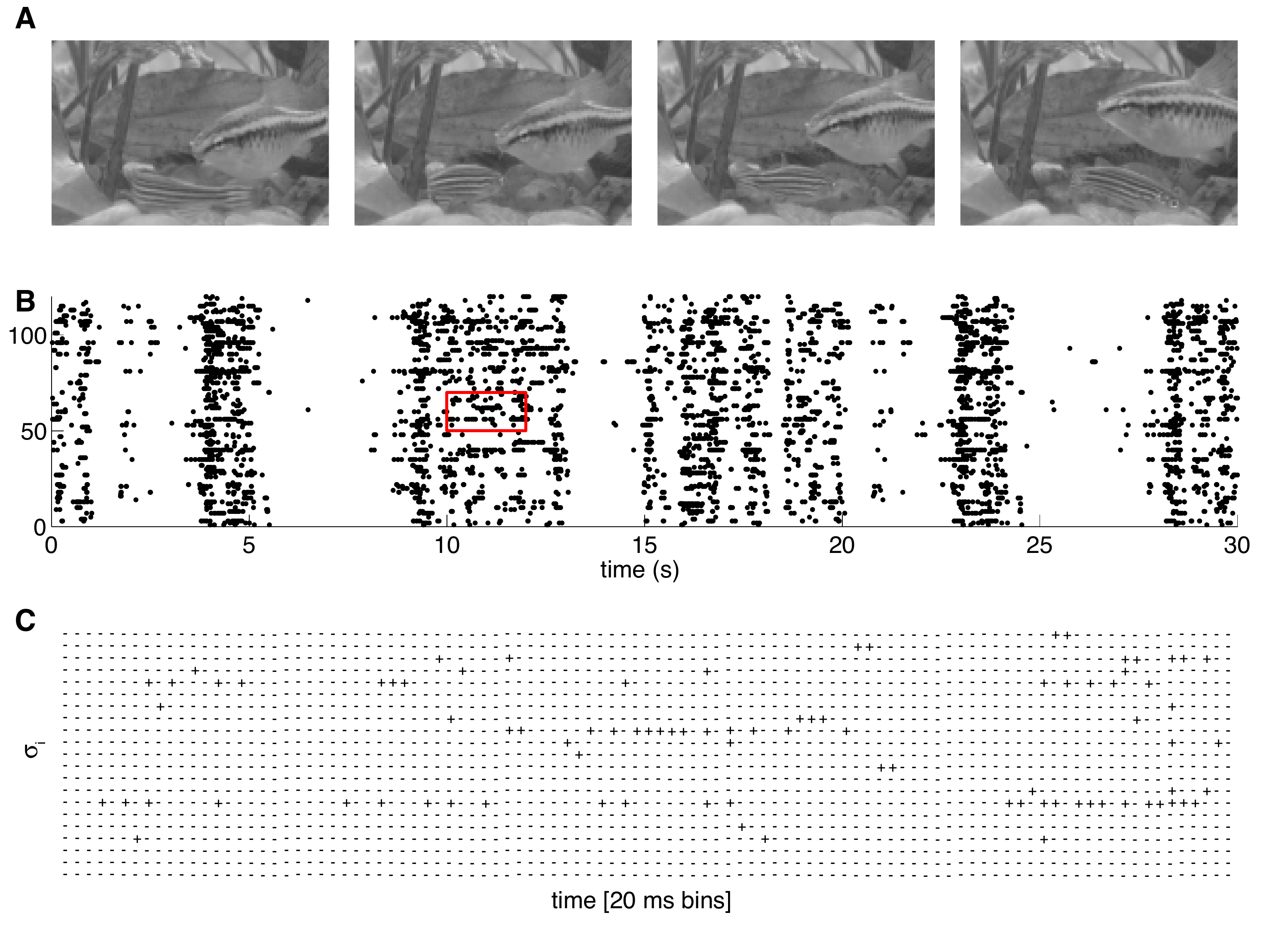} 
\caption{{\bf A schematic of the experiment.} {\bf (A)} Four frames from the natural movie stimulus showing swimming fish and water plants. {\bf (B)} The responses of a set of 120 neurons to a single stimulus repeat, black dots designate spikes. {\bf (C)} The raster for a zoomed-in region designated by a red square in (B), showing the responses discretized into $\Delta \tau=20\e{ms}$ time bins, where $\sigma_{\rm i} = -1$ represents a silence (absence of spike) of neuron ${\rm i}$, and $\sigma_{\rm i}=+1$ represents a spike.  \label{f1}}
\end{figure}

Here we return to the salamander retina, in experiments that exploit a new generation of multi--electrode arrays and associated spike--sorting algorithms \cite{marre+al_12}.  As schematized in Fig~\ref{f1}, these methods make  it possible to record from $N=100-200$ ganglion cells in the relevant densely interconnected patch, while projecting natural movies onto the retina.  Access to these large populations poses new problems for the inference of maximum entropy models, both in principle and in practice.  What we find is that, with extensions of algorithms developed previously \cite{Broderick+al_08}, it is possible to infer maximum entropy models for more than one hundred neurons, and that with nearly two hours of data there are no signs of ``overfitting.''    We have built models that match the mean probability of spiking for individual neurons, the correlations between spiking in pairs of neurons, and the distribution of summed activity in the network (i.e., the probability that $K$ out of the $N$ neurons spike in the same small window of time \cite{rolls,simplest,okun}).  We will see that models which satisfy all these experimental constraints provide a strikingly accurate description of the states taken on by the network as a whole, that these states {\em are} collective, and that the collective behavior predicted by our models has implications for how the retina encodes visual information.

\section{Maximum entropy}

The idea of maximizing entropy has its origin in thermodynamics and statistical mechanics.  The idea that we can use this principle to build models of systems that are not in thermal equilibrium is more recent, but still more than fifty years old \cite{jaynes_57}.  Here we provide a description of this approach which we hope makes the ideas accessible to a broad audience.

We imagine a neural system exposed to a stationary stimulus ensemble, in which simultaneous recordings of $N$ neurons can be made.  In small windows of time,  as we see in Fig~\ref{f1}, a single neuron ${\rm i}$ either does ($\sigma_{\rm i}=+1$) or does not ($\sigma_{\rm i} = -1$) generate an action potential or spike  \cite{spikes}; the state of the entire network in that time bin is therefore described by a ``binary word''  
$\{\sigma_{\rm i}\}$.  As the system responds to its inputs, it visits each of these states with some probability $P_{\rm expt}(\{\sigma_{\rm i}\})$.  Even before we ask what the different states mean, for example as codewords in a representation of the sensory world, specifying  this distribution requires us to determine the probability of each of $2^N$ possible states.  Once   $N$ increases beyond $\sim 20$, brute force sampling from data is no longer a general strategy for ``measuring'' the underlying distribution. 

Even when there are many, many possible states of the network, experiments of reasonable size  can be sufficient to estimate the averages or expectation  values of various functions of the state of the system,  $\langle f_\mu(\{\sigma_{\rm i}\})\rangle_{\rm expt} $ , where the averages are taken across data collected over the course of the experiment.  
The goal of the maximum entropy construction is to search for the probability distribution $P^{(\{f_\mu\})}(\{\sigma_{\rm i}\})$ that matches these experimental measurements but otherwise is as unstructured as possible. Minimizing structure means maximizing entropy \cite{jaynes_57}, and for any set of moments or statistics that we want to match, the form of the maximum entropy distribution can be found analytically:
\begin{eqnarray}
P^{(\{f_\mu\})}(\{\sigma_{\rm i}\}) &=& \frac{1}{Z(\{g_\mu\})}\exp\left({-\mathcal{H}}\right) \label{boltz}\\
\mathcal{H}(\{\sigma_{\rm i}\}) &=&-\sum_{\mu=1}^L g_\mu f_\mu(\{\sigma_{\rm i}\}), \label{ham}\\
Z(\{g_\mu\}) &=& \sum_{\{\sigma_{\rm i}\}} \exp\left({-\mathcal{H}}\right) ,
\end{eqnarray}
where $\mathcal{H}(\{\sigma_{\rm i}\})$ is the effective ``energy'' function or the Hamiltonian of the system, and the partition function $Z(\{g_\mu\})$ ensures that the distribution is normalized. The couplings $g_\mu$ must be set such that the expectation values of all constraint functions $\{\langle f_\mu\rangle_P\}$, $\mu=1,\dots,L$,  over the distribution $P$ match those measured in the experiment:
\begin{equation}
\langle f_\mu\rangle_{P} \equiv \sum_{\{\sigma_{\rm i}\}} f_\mu(\{\sigma_{\rm i}\})
P(\{\sigma_{\rm i}\}) =\frac{\partial \log Z}{\partial g_\mu}=\langle f_\mu\rangle_{\rm expt}. \label{con1}
\end{equation}
These equations might be hard to solve,  but they are  guaranteed to have exactly one solution for the  couplings $g_\mu$ given any set of measured expectation values \cite{kkt}.

Why should we study the neural vocabulary, $P(\{\sigma_{\rm i}\})$, at all? In much previous work on neural coding, the focus has been on constructing models for a ``codebook'' which can predict the response of the neurons to arbitrary stimuli, $P(\{\sigma_{\rm i}\}|{\rm stimulus})$ \cite{pillow,sdme}, or on building a ``dictionary'' that describes the stimuli consistent with particular patterns of activity, $P({\rm stimulus}| \{\sigma_{\rm i}\})$ \cite{spikes}.  In a natural setting, stimuli are drawn from a space of very high dimensionality,  so constructing these ``encoding'' and ``decoding'' mappings between the stimuli and responses is very challenging and often involves making strong assumptions  about how stimuli drive neural spiking (e.g. through linear filtering of the stimulus) \cite{fairhall_berry,keat,pillow,kolia}.  By trying to understand directly the total distribution of responses, $P(\{\sigma_{\rm i}\})$, rather than the conditional distribution, $P(\{\sigma_{\rm i}\}|{\rm stimulus})$, we take a very different approach.

Already when we study the smallest possible network, i.e. a pair of interacting neurons, the usual approach is to measure the correlation between spikes generated in the two cells, and to dissect this correlation into contributions which are intrinsic to the network and those which are ascribed to common, stimulus driven inputs.  The idea of decomposing correlations  dates back to a time when it was hoped that correlations among spikes could be used to map the synaptic connections between neurons \cite{perkel+bullock_68}.  In fact, in a highly interconnected system, the dominant source of correlations between two neurons---even if they are entirely intrinsic to the network---will always be through the multitude of indirect paths involving other neurons \cite{ginzburg+sompolinsky_94}.  Regardless of the source of these correlations, however, the question of whether they are driven by the stimulus or are intrinsic to the network is not a question that the brain can answer.  We, as external observers, can repeat the stimulus exactly, and search for correlations conditional on the stimulus, but this is not accessible to the organism.  The brain has access only to the output of the retina: the patterns of activity which are drawn from the distribution $P(\{\sigma_{\rm i}\})$.   If the responses $\{\sigma_{\rm i}\}$ are codewords for the visual stimulus, then the entropy of this distribution sets the capacity of the code to carry information.  Word by word, $-\log P(\{\sigma_{\rm i}\})$ determines how surprised the brain should be by each particular pattern of response, including the possibility that the response was corrupted by noise in the retinal circuit and thus should be corrected or ignored \cite{hopfield_08}.  In a very real sense, what the brain ``sees'' are sequences of states drawn from  $P(\{\sigma_{\rm i}\})$.  In the same spirit that 
many groups have studied the statistical structures of natural scenes \cite{field, dong+atick, ruderman,simoncelli+schwartz,bethge,stephens}, we would like to understand the statistical structure of the codewords that represent these scenes.

The maximum entropy method is not a model for network activity.  Rather it is a framework for building models, and to implement this framework we have to choose which functions of the network state $f_\mu(\{\sigma_{\rm i}\})$ we think are interesting.  The hope is that while there are $2^N$ states of the system as a whole, there is a much smaller number of measurements, $\{f_\mu (\{\sigma_{\rm i}\})\}$, with $\mu = 1 , 2, \cdots , L$ and $L \ll 2^N$, which will be sufficient to capture the essential structure of the collective behavior in the system.  We emphasize that this is a hypothesis, and must be tested.  How should we choose the functions $f_\mu (\{\sigma_{\rm i}\})$?  In this work we consider three classes of possibilities:

(A) We expect that networks have very different behaviors depending on the overall probability that neurons generate spikes as opposed to remaining silent.  Thus, our first choice of functions to constrain in our models is the set of mean spike probabilities or firing rates, which is equivalent to constraining $\langle \sigma_{\rm i}\rangle$, for each neuron $\rm i$.  These constraints contribute a term to the energy function
\begin{equation}
{\mathcal H}^{(1)} = - \sum_{{\rm i}=1}^N h_{\rm i} \sigma_{\rm i} .
\label{HA}
\end{equation}
Note that $\langle \sigma_{\rm i}\rangle = -1 + 2\bar r_{\rm i} \Delta\tau$, where $\bar r_{\rm i}$ is the mean spike rate of neuron $\rm i$, and $\Delta\tau$ is the size of the time slices that we use in our analysis, as in Fig~\ref{f1}.   \emph{Maximum entropy models that constrain only the firing rates of all the neurons (i.e. $\mathcal{H}=\mathcal{H}^{(1)}$) are called ``independent models''; we denote their  distribution functions by $P^{(1)}$.}

(B) As a second constraint we take the correlations between neurons, two by two.  This corresponds to measuring 
\begin{equation}
C_{\rm ij} = \langle \sigma_{\rm i}\sigma_{\rm j}\rangle-\langle\sigma_{\rm i}\rangle\langle\sigma_{\rm j}\rangle \label{def_Cij}
\end{equation}
for every pair of cells $\rm ij$. These constraints contribute a term to the energy function
\begin{equation}
{\mathcal H}^{(2)} = - {1\over 2} \sum_{{\rm i}, {\rm j} =1}^N J_{\rm ij} \sigma_{\rm i} \sigma_{\rm j}.
\label{HB}
\end{equation}
It is more conventional to think about correlations between two neurons in terms of their spike trains.  If we define
\begin{equation}
\rho_{\rm i}(t) = \sum_n \delta (t - t_n^{\rm i}) ,
\end{equation}
where neuron $\rm i$ spikes at times $t_n^{\rm i}$, then the spike--spike correlation function is \cite{spikes}
\begin{equation}
C_{\rm ij}^{\rm spike} (t-t') = \langle \rho_{\rm i}(t ) \rho_{\rm j}(t')\rangle - \langle \rho_{\rm i}\rangle \langle \rho_{\rm j}\rangle  ,
\end{equation}
and we also have the average spike rates $\bar r_{\rm i} = \langle \rho_{\rm i}\rangle$.  The correlations among the discrete spike/silence variables $\sigma_{\rm i}, \sigma_{\rm j}$ then can be written as
\begin{equation}
C_{\rm ij} = 4\int_0^{\Delta\tau} dt \int_0^{\Delta\tau} dt' C_{\rm ij}^{\rm spike} (t-t') .
\end{equation}
\emph{Maximum entropy models that constrain average firing rates and correlations (i.e. $\mathcal{H}=\mathcal{H}^{(1)}+\mathcal{H}^{(2)}$) are called ``pairwise models''; we denote their distribution functions by $P^{(1,2)}$.}

(C) Firing rates and pairwise correlations focus on the properties of particular neurons.  As an alternative, we can consider quantities that refer to the network as a whole, independent of the identity of the individual neurons.  A simple example is the ``distribution of synchrony'' (also called ``population firing rate''), that is the probability $P_N(K)$ that $K$ out of the $N$ neurons spike in the same small slice of time.    We can count the number of neurons that spike by summing all of the $\sigma_{\rm i}$, remembering that we have $\sigma_{\rm i} =1$ for spikes and $\sigma_{\rm i} = -1$ for silences.  Then 
\begin{equation}
P_N(K) = {\bigg\langle}\delta\left( \sum_{{\rm i}=1}^N \sigma_{\rm i} , 2K-N\right){\bigg\rangle} ,
\end{equation}
where
\begin{eqnarray}
\delta\left(n,n\right) &=& 1;\\
\delta\left(n,m\neq n\right) &=& 0 .
\end{eqnarray}
If we know the distribution $P_N(K)$, then we know all its moments, and hence we can think of the functions  $f_\mu (\{\sigma_{\rm i}\})$ that we are constraining as being
\begin{eqnarray}
f_1 (\{\sigma_{\rm i}\}) &=&  \sum_{{\rm i}=1}^N \sigma_{\rm i} ,\\
f_2 (\{\sigma_{\rm i}\}) &=& \left( \sum_{{\rm i}=1}^N \sigma_{\rm i}\right)^2 ,\\
f_3 (\{\sigma_{\rm i}\}) &=& \left( \sum_{{\rm i}=1}^N \sigma_{\rm i}\right)^3 ,
\end{eqnarray}
and so on.  Because there are only $N$ neurons, there are only $N+1$ possible values of $K$, and hence only $N$ unique moments.  Constraining all of these moments contributes a term to the energy function
\begin{equation}
{\mathcal H}^{(K)} = -\sum_{K=1}^N \lambda_K \left( \sum_{{\rm i}=1}^N \sigma_{\rm i}\right)^K =  -V \left(\sum_{{\rm i}=1}^N \sigma_{\rm i} \right) ,
\label{HC}
\end{equation}
where $V$ is an effective potential \cite{simplest,okun}.  \emph{Maximum entropy models that constrain average firing rates, correlations, and the distribution of synchrony (i.e. $\mathcal{H}=\mathcal{H}^{(1)}+\mathcal{H}^{(2)}+\mathcal{H}^{(K)}$) are called ``K-pairwise models''; we denote their distribution functions by $P^{(1,2,K)}$.}

It is important that the mapping between maximum entropy models and a Boltzmann distribution with some effective energy function is {\em not} an analogy, but rather a mathematical equivalence.  In using the maximum entropy approach we are {\em not} assuming that the system of interest is in some thermal equilibrium state (note that there is no explicit temperature in Eq~(\ref{boltz})), nor are we assuming that there is some mysterious force which drives the system to a state of maximum entropy.  We are also not assuming that the temporal dynamics of the network is described by Newton's laws or Brownian motion on the energy landscape.  What we are doing is making models that are consistent with certain measured quantities, but otherwise have as little structure as possible.  As noted above, this is the opposite of what we usually do in building models or theories---rather than trying to impose some hypothesized structure on the world, we are trying to remove all structures that are not strictly required by the data.

The mapping to a Boltzmann distribution is not an analogy, but if we take the energy function more literally we are making use of analogies.  Thus, the term ${\mathcal H}^{(1)}$ that emerges from constraining the mean spike probabilities of every neuron is analogous to a magnetic field being applied to each spin, where spin ``up'' ($\sigma_{\rm i} = +1$) marks a spike and spin ``down'' ($\sigma_{\rm i} = -1$) denotes silence.  Similarly, the term ${\mathcal H}^{(2)}$ that emerges from constraining the pairwise correlations among neurons corresponds to a ``spin--spin'' interaction which tends to favor neurons firing together ($J_{\rm ij} >0$) or not ($J_{\rm ij} < 0$).  Finally, the constraint on the overall distribution of activity generates a term ${\mathcal H}^{(K)}$ which we can interpret as resulting from the interaction between all the spins/neurons in the system and one other, hidden degree of freedom, such as an inhibitory interneuron.  These analogies can be useful, but need not be taken literally.

\section{Can we learn the model?}

We have applied the maximum entropy framework to the analysis of one large experimental data set on the responses of ganglion cells in the salamander retina to a repeated, naturalistic movie.   These data are collected using a new generation of multi--electrode arrays that allow us to record from a large fraction of the neurons in a $450\times 450\e{\mu m}$  patch, which contains a total of $\sim 200$ ganglion cells \cite{marre+al_12}, as in Fig~\ref{f1}.  In the present data set, we have selected 160 neurons that pass standard tests for the stability of spike waveforms, the lack of refractory period violations, and the stability of firing across the duration of the experiment (see Methods and Ref \cite{marre+al_12}).  The visual stimulus is  a greyscale movie of swimming fish and swaying water plants in a tank; the analyzed chunk of movie is $19\,{\rm s}$ long, and the recording was stable through 297 repeats, for a total of more than $1.5\,{\rm hrs}$ of data. As has been found in previous experiments in the retinas of multiple species \cite{mastronarde,devries,brivanlou,torng,schneidman+al_06}, we found that correlations among neurons are most prominent on the $\sim 20\,{\rm ms}$ time scale, and so we chose to discretize the spike train into $\Delta\tau = 20\,{\rm ms}$ bins.

 \begin{figure}
\includegraphics[width=\linewidth]{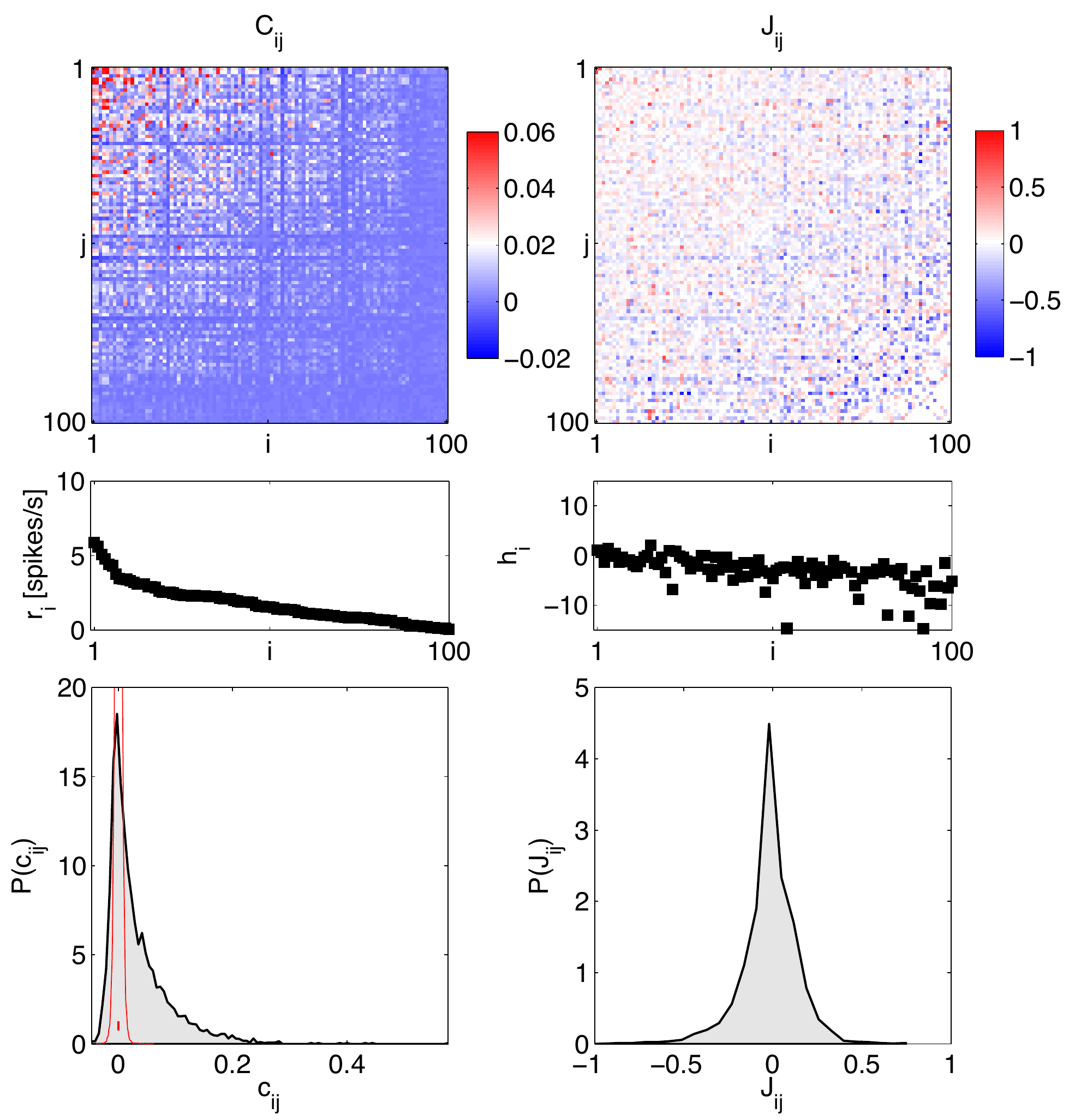} 
\caption{{\bf Learning the pairwise maximum entropy model  for a 100 neuron subset.} A subgroup of 100 neurons from our set of 160 has been sorted by the firing rate.  At left, the statistics of the neural activity: correlations $C_{\rm ij}=\langle \sigma_{\rm i}\sigma_{\rm j}\rangle-\langle\sigma_{\rm i}\rangle\langle\sigma_{\rm j}\rangle$ (top), firing rates (equivalent to $\langle \sigma_{\rm i}\rangle$; middle), and the distribution of correlation coefficients $c_{\rm ij}$ (bottom).  The red distribution is the distribution of differences between two halves of the experiment, and the small red error bar marks the standard deviation of correlation coefficients in fully shuffled data ($1.8\times 10^{-3}$). 
At right, the parameters of a pairwise maximum entropy model [${\mathcal H}$ from  Eq~(\ref{Hpair})] that reproduces these data:  coupling constants $J_{\rm ij}$ (top),  fields $h_{\rm i}$ (middle), and the distribution of couplings in this group of neurons.  \label{f2}}.
\end{figure}

Maximum entropy models have a simple form [Eq~(\ref{boltz})] that connects precisely with statistical physics.  But to complete the construction of a maximum entropy model, we need to impose the condition that averages in the maximum entropy distribution match the experimental measurements, as in Eq~(\ref{con1}).  This amounts to finding all the coupling constants $\{g_\mu\}$ in Eq~(\ref{ham}).  This is, in general, a hard problem.  We need not only to solve this problem, but also convince ourselves that our solution is meaningful, and that it does not reflect overfitting to the limited set of data at our disposal.   A detailed account of the numerical solution to this inverse problem is given in Appendix~\ref{App_EntEst}.

In Fig~\ref{f2} we show an example of $N=100$ neurons from a small patch of the salamander retina, responding to naturalistic movies.  We notice that correlations are weak, but widespread, as in previous experiments on smaller groups of neurons \cite{schneidman+al_06,tkacik+al_06,tkacik+al_09,puchalla+al_05,segev_physiol}.  Because the data set is very large, the threshold for reliable detection of correlations is very low; if we shuffle the data completely by permuting time and repeat indices independently for each neuron, the standard deviation of correlation coefficients,
\begin{equation}
c_{\rm ij} = {{\langle \sigma_{\rm i}\sigma_{\rm j}\rangle-\langle\sigma_{\rm i}\rangle\langle\sigma_{\rm j}\rangle}\over{\sqrt{(1-\langle \sigma_{\rm i}\rangle^2)(1-\langle\sigma_{\rm j}\rangle^2)}}},
\end{equation}
is $\sigma_c = 1.8\times 10^{-3}$, as shown in Fig~\ref{f2}C, vastly smaller than the typical correlations that we observe (median $1.7\E{-2}$, 90\% of values between $-1.6\E{-2}$ and $1.37\E{-1}$).  More subtly, this means that only a few percent of the correlation coefficients are within error bars of zero, and there is no sign that there is a finite fraction of pairs that have truly zero correlation---the distribution of correlations across the population seems continuous. Note that, as customary, we report normalized correlation coefficients ($c_{\rm ij}$, between -1 and 1), while maximum entropy formally constrains an equivalent set of unnormalized second order moments, $C_{\rm ij}$ [Eq (\ref{def_Cij})].

 \begin{figure}[t] 
\includegraphics[width=\linewidth]{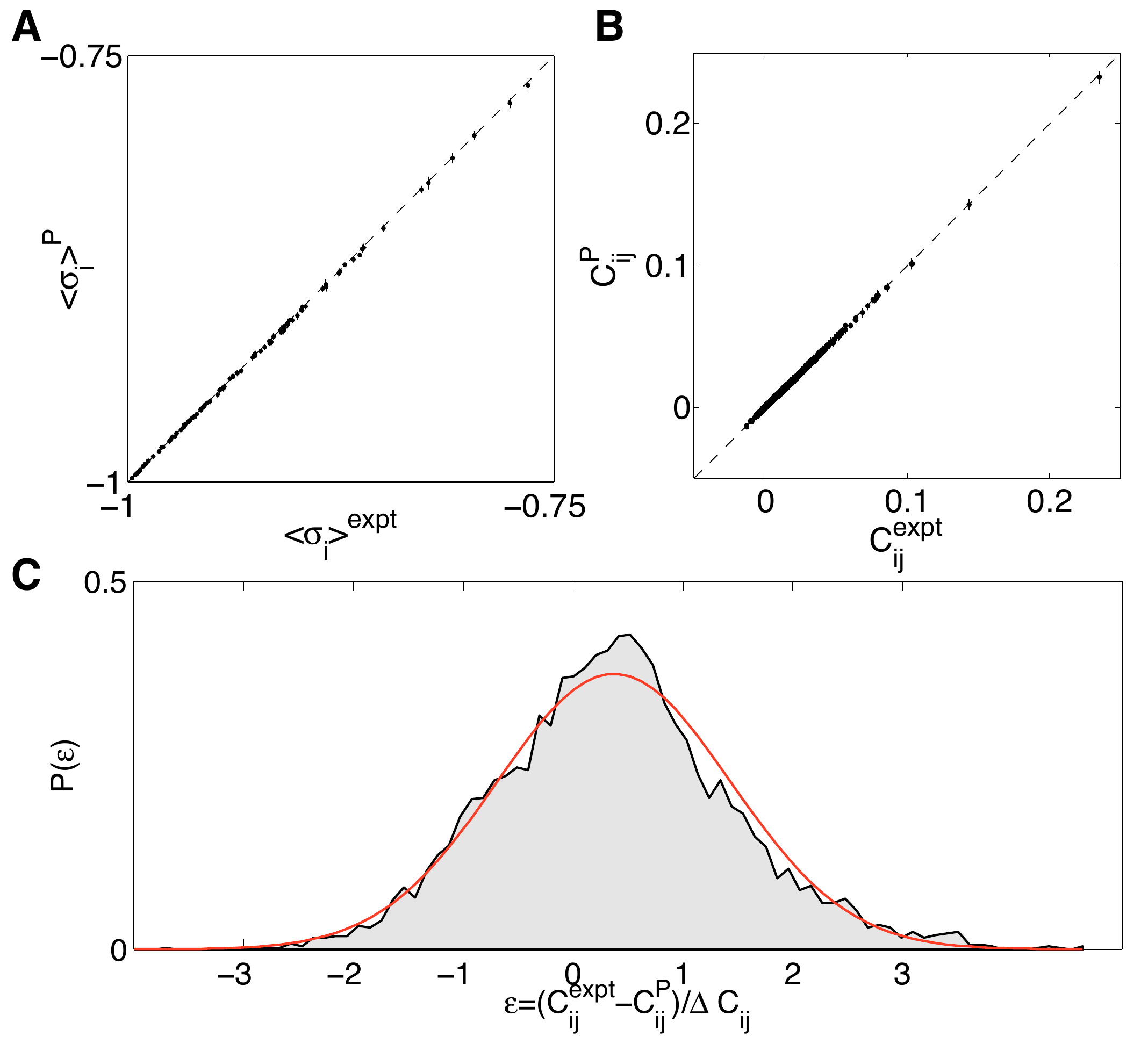} 
\caption{{\bf Reconstruction precision  for a 100 neuron subset.} Given the reconstructed Hamiltonian of the pairwise model, we used an independent Metropolis Monte Carlo (MC) sampler to assess how well the constrained model statistics (mean firing rates {\bf (A)}, covariances {\bf (B)}, plotted on y-axes) match the measured statistics (corresponding x-axes). Error bars on data computed by bootstrapping; error bars on MC estimates obtained by repeated MC runs generating a number of samples that is equal to the original data size. {\bf (C)} The distribution of the difference between true and model values for $\sim 5\E{3}$ covariance matrix elements, normalized by the estimated error bar in the data; red overlay is a Gaussian with zero mean and unit variance. The distribution has nearly Gaussian shape with a width of $\approx 1.1$, showing that the learning algorithm reconstructs the covariance statistics to within measurement precision.  \label{f3}}
\end{figure}

We began by constructing maximum entropy models that match the mean spike rates and pairwise correlations, i.e. ``pairwise models,'' whose distribution is, from Eqs~(\ref{HA},\,\ref{HB}),
\begin{eqnarray}
P^{(1,2)}(\{\sigma_{\rm i}\}) &=& {1\over Z} \exp\left[ -{\mathcal H} (\{\sigma_{\rm i}\})\right]\nonumber\\
{\mathcal H} &=& - \sum_{{\rm i}=1}^N h_{\rm i} \sigma_{\rm i} - {1\over 2} \sum_{{\rm i}, {\rm j} =1}^N J_{\rm ij} \sigma_{\rm i} \sigma_{\rm j} .
\label{Hpair}
\end{eqnarray}
When we reconstruct the coupling constants of the maximum entropy model, we see that the ``interactions'' $J_{\rm ij}$ among neurons are widespread, and almost symmetrically divided between positive and negative values; for more details see Appendix~\ref{App_EntEst}.
 Figure~\ref{f3} shows that the model we construct really does satisfy the constraints, so that the differences, for example, between the measured and predicted correlations among pairs of neurons are within the experimental errors in the measurements.

With $N=100$ neurons, measuring the mean spike probabilities and all the pairwise correlations means that we estimate $N(N+1)/2 = 5050$ separate quantities.  This is a  large number, and it is not clear that we are safe in taking all these measurements at face value.  It is possible, for example, that with a finite data set the errors in the different elements of the correlation matrix $C_{\rm ij}$ are sufficiently strongly correlated that we don't really know the matrix as a whole with high precision, even though the individual elements are measured very accurately.    This  is a question about overfitting:  is it possible that the parameters $\{h_{\rm i},J_{\rm ij}\}$ are being finely tuned to match even the statistical errors in our data?  

 \begin{figure}[b]
\includegraphics[width=\linewidth]{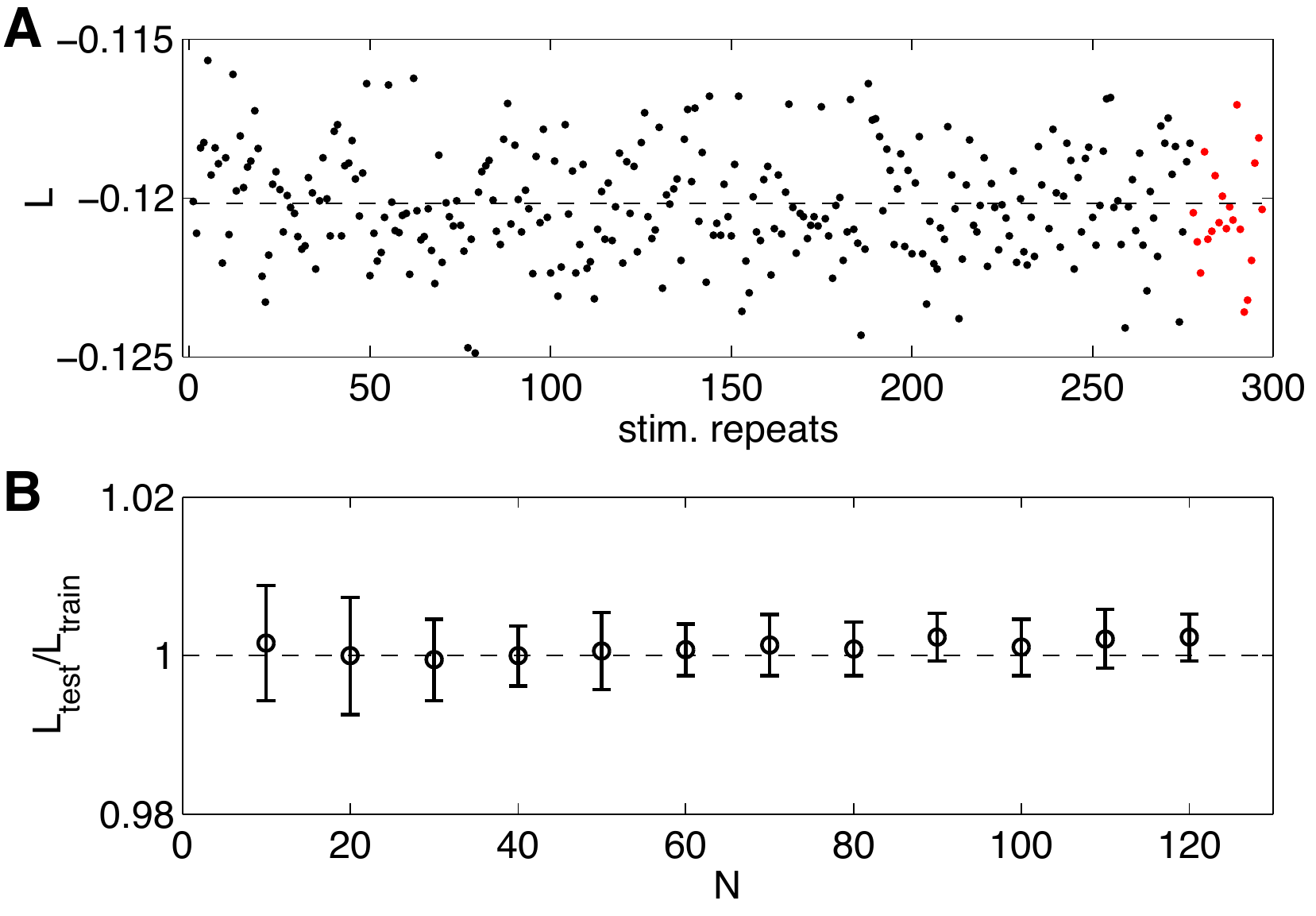} 
\caption{{\bf A test for overfitting.} {\bf (A)} The per-neuron average log-probability of data (log-likelihood, $L=\langle \log P(\sigma)\rangle_{\rm expt}/N$) under the pairwise model of Eq~(\ref{Hpair}), computed on the training repeats (black dots) and on the testing repeats (red dots),  for the same  group of $N=100$ neurons shown in Fig~\ref{f2} and \ref{f3}. Here the repeats have been reordered so that the training repeats precede testing repeats; in fact, the choice of test repeats is random. {\bf (B)} The ratio of the log-likelihoods on test vs training data, shown as a function of the network size $N$.  Error bars are the standard deviation across 30 subgroups at each value of $N$. \label{f4}}
\end{figure}

To test for overfitting (Fig \ref{f4}), we exploit the fact that the stimuli consist of a short movie repeated many times.  We can choose a random $90\%$ of these repeats from which to learn the parameters of the maximum entropy model, and then check that the probability of the data in the other $10\%$ of the experiment is predicted to be the same, within errors.  We see in Fig~\ref{f4} that this is true, and that it remains true as we expand from $N=10$ neurons (for which we surely have enough data) out to $N=120$, where we might have started to worry.  Taken together, Figs~\ref{f2}, \ref{f3}, and \ref{f4} suggest strongly that our data and algorithms are sufficient to construct maximum entropy models, reliably, for networks of more than one hundred neurons.

\section{Do the models work?}

How well do our maximum entropy models describe the behavior of large networks of neurons? The models predict the probability of occurrence for all possible combinations of spiking and silence in the network, and it seems natural to use this huge predictive power to test the models.  In small networks, this is a useful approach.  Indeed, much of the interest in the maximum entropy approach derives from the success of models based on mean spike rates and pairwise correlations, as in Eq~(\ref{Hpair}), in reproducing the probability distribution over states in networks of size $N=10-15$ \cite{schneidman+al_06,shlens+al_06}.  With $N=10$, there are $2^{10} =1024$ possible combinations of spiking and silence, and reasonable experiments are sufficiently long to estimate the probabilities of all of these individual states.  But with $N=100$, there are $2^{100}\sim 10^{30}$ possible states, and so it is not possible to ``just measure'' all the probabilities.  Thus, we need another strategy for testing our models.

\begin{figure}[t]
\includegraphics[width=\linewidth]{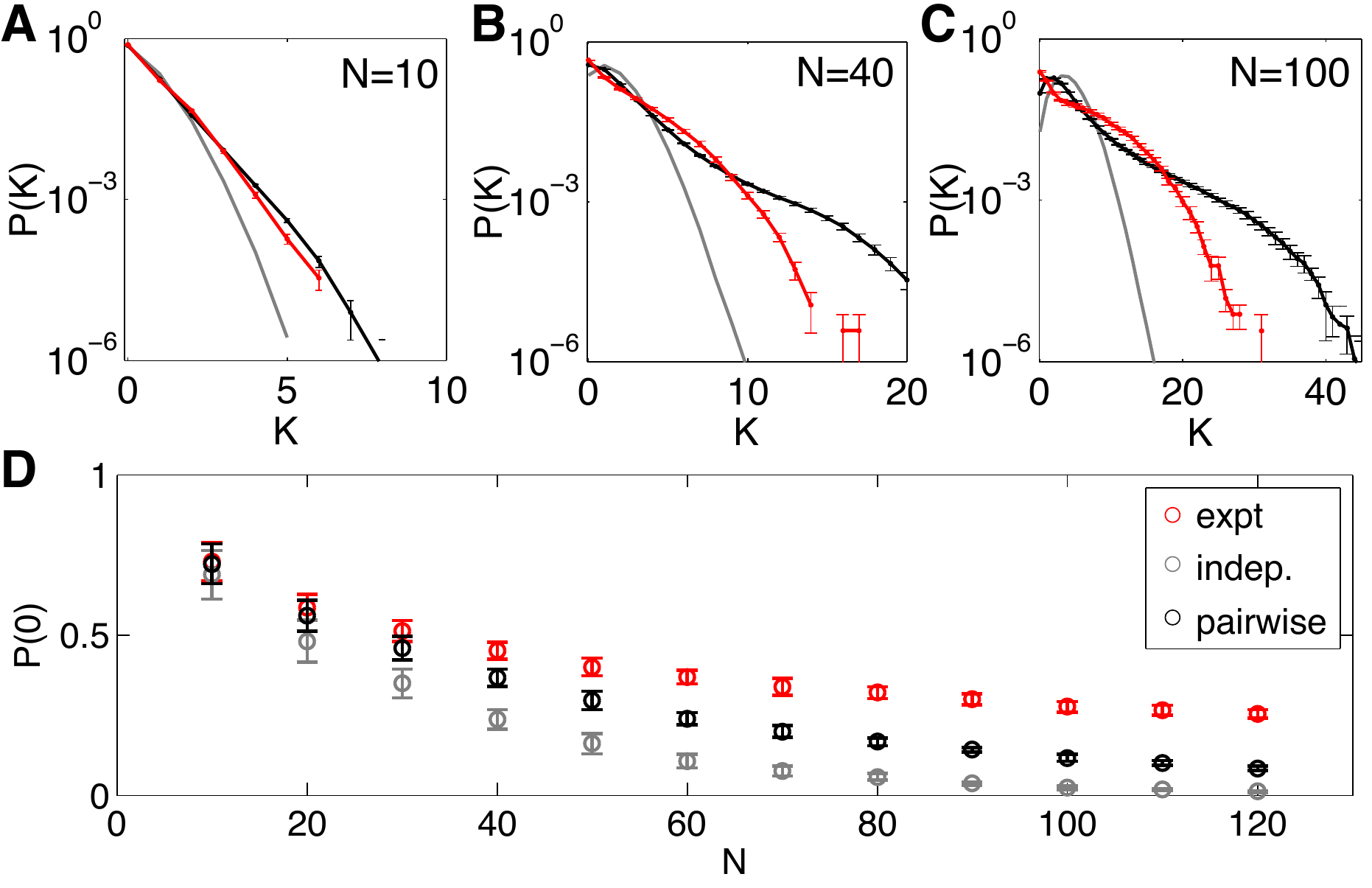} 
\caption{{\bf Predicted vs measured probability of $K$ simultaneous spikes  (spike synchrony).} {\bf (A-C)} $P_N(K)$ for subnetworks of size $N=10, 40, 100$; error bars are s.d. across random halves of the duration of the experiment. For $N=10$ we already see large deviations from an independent model, but these are captured by the pairwise model.  At $N=40$ (B), the pairwise models miss the tail of the distribution, where $P(K)< 10^{-3}$.  At $N=100$ (C), the deviations between the pairwise model and the data are more substantial. {\bf (D)} The probability of silence in the network, as a function of population size; error bars are s.d. across 30 subgroups of a given size $N$. Throughout, red shows the data, grey the independent model, and black the pairwise model.\label{f5}}
\end{figure}

Striking (and model--independent) evidence for nontrivial collective behavior in these networks is obtained by asking for the probability that $K$ out of the $N$ neurons generate a spike in the same small window of time, as shown in Fig~\ref{f5}.  This distribution, $P_N(K)$, should become Gaussian at large $N$ if the neurons are independent, or nearly so, and we have noted that the correlations between pairs of cells are weak.  Thus $P_2(K)$ is very well approximated by an independent model, with fractional errors on the order of the correlation coefficients, typically less than $\sim 10\%$.  But, even in groups of $N=10$ cells, there are substantial departures from the predictions of an independent model (Fig \ref{f5}A). In groups of $N=40$ cells, we see $K=10$ cells spiking synchronously with probability $\sim 10^4$ times larger than expected from an independent model (Fig \ref{f5}B), and the departure from independence is even larger at $N=100$ (Fig~\ref{f5}C).

Maximum entropy models that match the mean spike rate and pairwise correlations in a network make an unambiguous, quantitative prediction for $P_N(K)$, with no adjustable parameters.  In smaller groups of neurons, certainly for $N=10$, this prediction is quite accurate, and  accounts for most of the difference between the data and the expectations from an independent model, as shown in Fig~\ref{f5}.  But even at $N=40$ we see small deviations between the data and the predictions of the pairwise model.  Because the silent state is highly probable, we can measure $P_N(K=0)$ very accurately, and the pairwise models make errors of nearly a factor of three at $N=100$.   These errors are negligible when compared to the many orders of magnitude differences from an independent model, but they are highly significant.  The pattern of errors also is important, since in the real networks silence persists as being highly probable even at $N=100$ (and beyond), which is surprising \cite{simplest}, and the pairwise model doesn't quite capture this.  

\begin{figure}[tb]
\includegraphics[width=\linewidth]{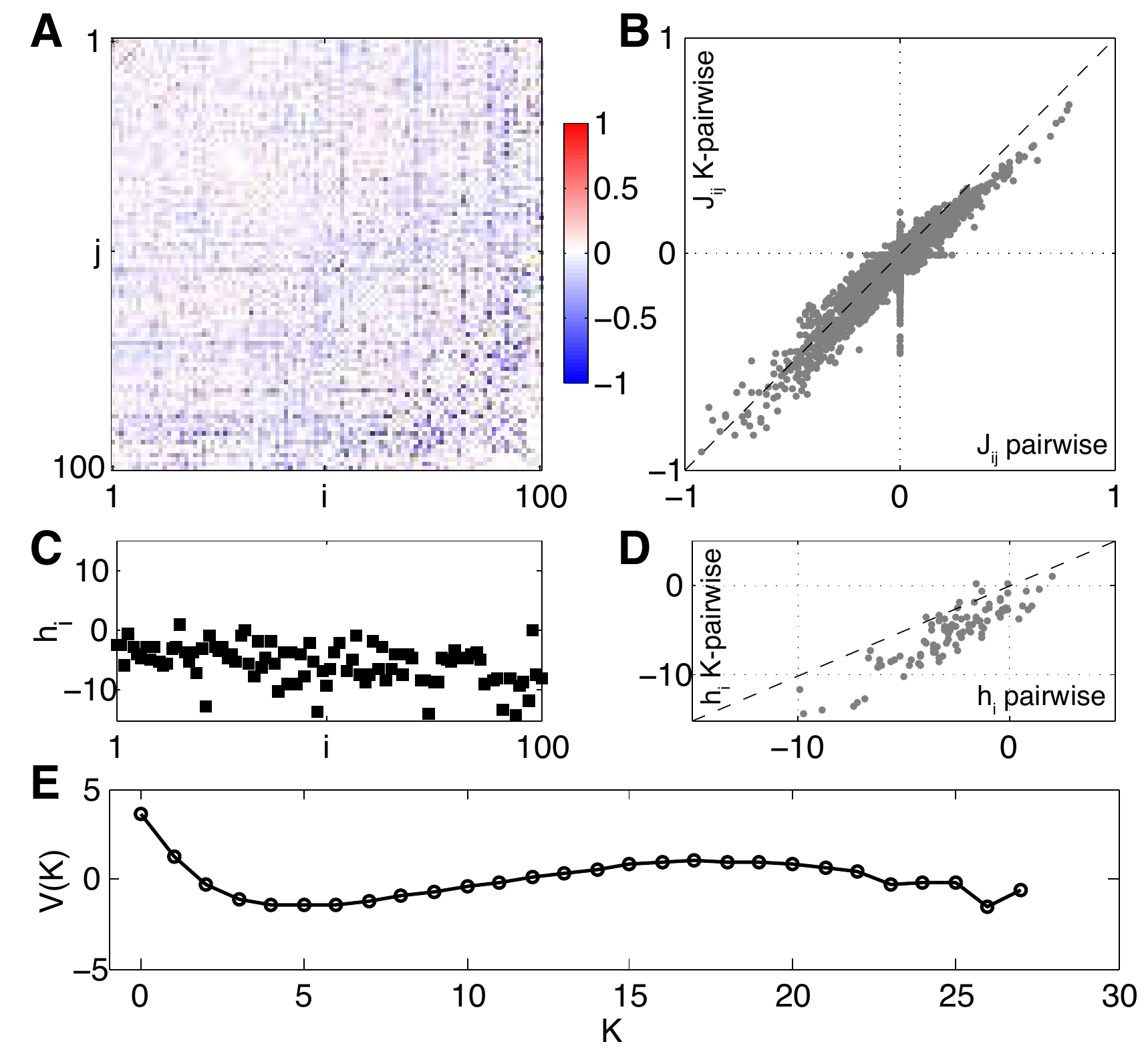} 
\caption{{\bf K-pairwise model for a the same group of $N=100$ cells shown in Fig~\ref{f1}.} The neurons are again sorted in the order of decreasing firing rates. {\bf (A)} Pairwise interactions, $J_{\rm ij}$, and the comparison with the interactions of the pairwise model, {\bf (B)}. {\bf (C)} Single-neuron fields, $h_{\rm i}$, and the comparison with the fields of the pairwise model, {\bf (D)}. {\bf (E)} The global potential, $V(K)$, where $K$ is the number of  synchronous spikes.   \label{f6}}
\end{figure}

If a model based on pairwise correlations doesn't quite account for the data, it is tempting to try and include correlations among triplets of neurons.  But at $N=100$ there are $N(N-1)(N-2)/6 \sim 1.6 \times 10^5$ of these triplets, so a model that includes these correlations is much more complex than one that stops with pairs.  An alternative is to use $P_N(K)$ itself as a constraint on our models, as explained above in relation to Eq~(\ref{HC}).  This defines the ``K-pairwise model,'' 
\begin{eqnarray}
P^{(1,2,K)}(\{\sigma_{\rm i}\}) &=& {1\over Z} \exp\left[ -{\mathcal H} (\{\sigma_{\rm i}\})\right]\label{HKpair} \\
{\mathcal H} (\{\sigma_{\rm i}\})&=& - \sum_{{\rm i}=1}^N h_{\rm i} \sigma_{\rm i} - {1\over 2} \sum_{{\rm i}, {\rm j} =1}^N J_{\rm ij} \sigma_{\rm i} \sigma_{\rm j}- V \left(\sum_{{\rm i}=1}^N \sigma_{\rm i} \right) ,  \nonumber
\end{eqnarray}
where the ``potential'' $V$ is chosen to match the observed distribution $P_N(K)$.  As noted above, we can think of this potential as providing a global regulation of the network activity, such as might be implemented by inhibitory interneurons with (near) global connectivity.  Whatever the mechanistic interpretation of this model, it is important that it is {\em not} much more complex than the pairwise model:  matching $P_N(K)$ adds only $\sim N$ parameters to our model, while the pairwise model already has $\sim N^2 /2$ parameters.  All of the tests given in the previous section can be redone in this case, and again we find that we can learn the K-pairwise models from the available data with no signs of overfitting. Figure~\ref{f6} shows the parameters of the K-pairwise model for the same group of $N=100$ neurons shown in Fig~\ref{f2}. Notice that the pairwise interaction terms $J_{\rm ij}$ remain roughly the same; the local fields $h_{\rm i}$ are also similar but have a shift towards more negative values.

\begin{figure}[b]
\includegraphics[width=0.8\linewidth]{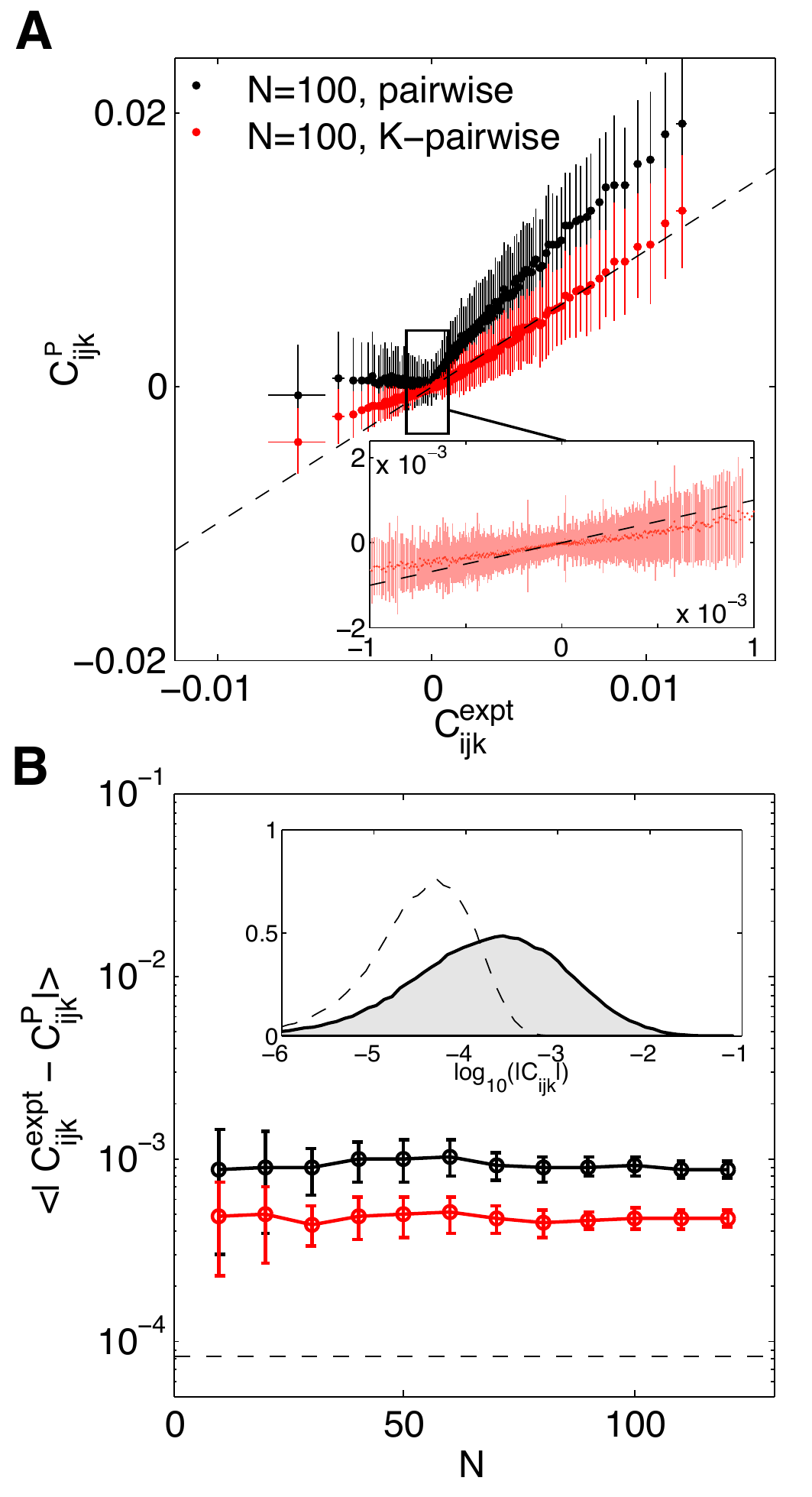} 
\caption{{\bf Predicted vs real connected three--point correlations, $C_{\rm ijk}$ from Eq (\ref{3pt}).} {\bf (A)} Measured $C_{\rm ijk}$  (x-axis) vs predicted by the model (y-axis), shown for an example 100 neuron subnetwork. The $\sim 1.6 \times 10^5$ triplets are binned into 1000 equally populated bins; error bars in x are s.d. across the bin. The corresponding values for the predictions are grouped together, yielding the mean and the s.d. of the prediction (y-axis). Inset shows a zoom-in of the central region, for the K-pairwise model.
{\bf (B)} Error in predicted three-point correlation functions as a function of subnetwork size $N$. Shown are mean absolute deviations of the model prediction from the data, for pairwise (black) and K-pairwise (red) models; error bars are s.d. across 30 subnetworks at each $N$, and the dashed line shows the mean absolute difference between two halves of the experiment. Inset shows the distribution of three--point correlations (grey filled region) and the distribution of differences between two halves of the experiment (dashed line); note the logarithmic scale.   \label{f7}}
\end{figure}

Since we didn't make explicit use of the triplet correlations in constructing the K-pairwise model, we can test the model by predicting these correlations.   In Fig~\ref{f7}A we show
\begin{equation}
C_{\rm ijk} \equiv {\bigg\langle} (\sigma_{\rm i} - \langle \sigma_{\rm i} \rangle ) (\sigma_{\rm j} - \langle \sigma_{\rm j} \rangle ) (\sigma_{\rm k} - \langle \sigma_{\rm k} \rangle ){\bigg\rangle}
\label{3pt}
\end{equation}
as computed from the real data and from the models, for a single group of $N=100$ neurons.  We see that pairwise models capture the rankings of the different triplets, so that more strongly correlated triplets are predicted to be more strongly correlated, but these models miss quantitatively, overestimating the positive correlations and failing to predict significantly negative correlations.  These systematic errors are largely corrected in the K-pairwise model, despite the fact that adding a constraint on $P_N(K)$ doesn't add any information about the identity of the neurons in the different triplets.   It is also interesting that this improvement in our predictions (as well as that in Fig~\ref{f8} below) occurs even though the numerical value of the effective potential $V_N(K)$ is quite small, as shown in Fig~\ref{f6}E.  Fixing the distribution of global activity thus seems to capture something about the network that individual spike probabilities and pairwise correlations have missed.

An interesting effect is shown in Fig~\ref{f7}B, where we look at the average absolute deviation between predicted and measured $C_{\rm ijk}$, as a function of the group size $N$. With increasing $N$ the ratio between the total number of (predicted)  three-point correlations and (fitted) model parameters is increasing (from $\approx 2$ at $N=10$ to $\approx 40$ for $N=120$), leading us to believe that predictions will grow progressively worse. Nevertheless, the average error in three-point prediction stays constant with network size, for both pairwise and K-pairwise models.  An attractive explanation is that, as $N$ increases, the models encompass larger and larger fractions of the interacting neural patch and thus decrease the effects of ``hidden'' units, neurons that are present but not included in the model; such unobserved units, even if they only interacted with other units in a pairwise fashion, could introduce effective higher-order interactions between observed units, thereby causing three-point correlation predictions to deviate from those of the pairwise model \cite{schneidman+al_03}.  The accuracy of the K-pairwise predictions is not quite as good as the errors in our measurements (dashed line in Fig~\ref{f7}B), but still very good, improving by a factor of $\sim 2$ relative to the pairwise model to well below $10^{-3}$.

\begin{figure}[b]
\includegraphics[width=\linewidth]{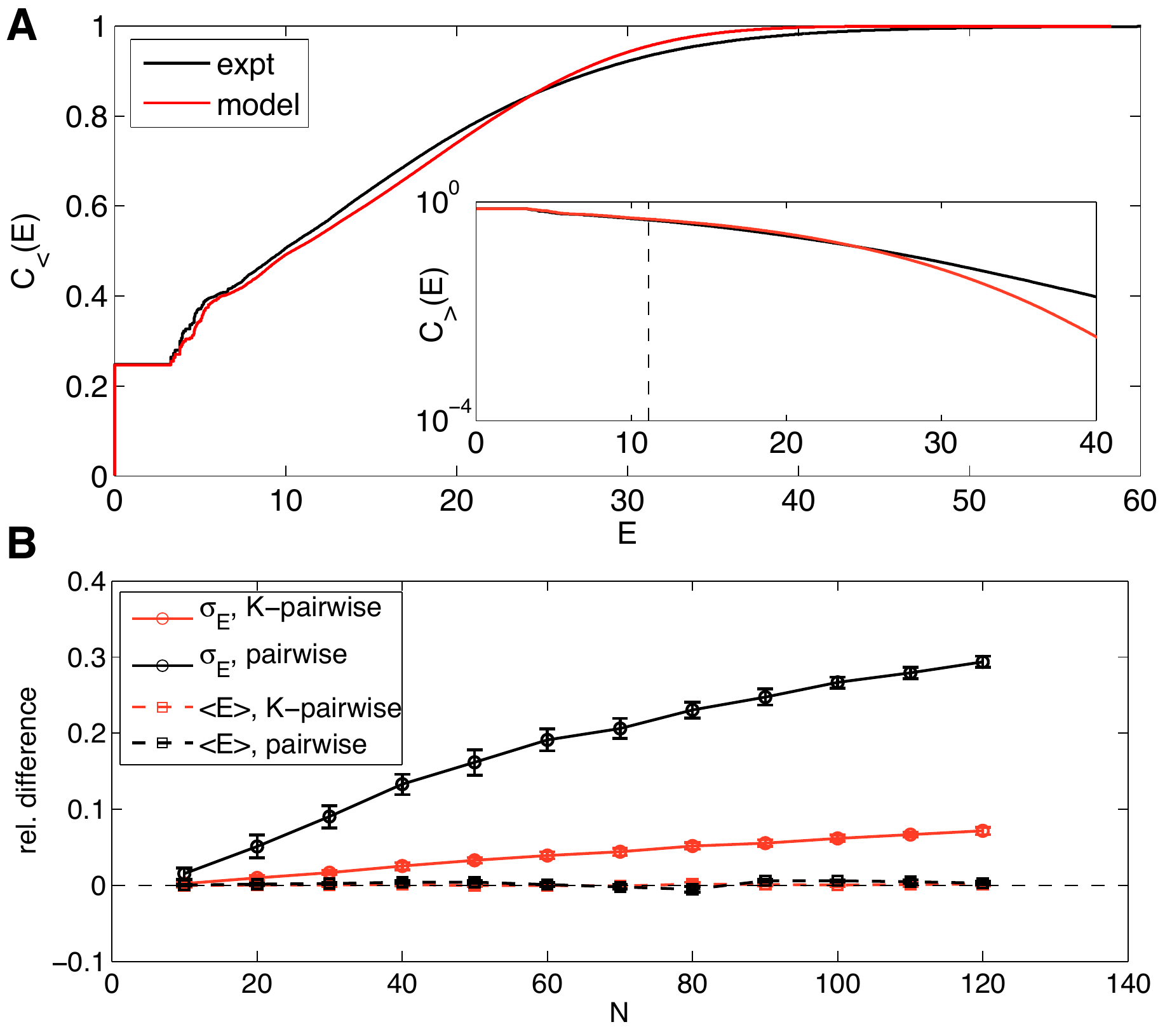} 
\caption{{\bf Predicted vs real distributions of energy, $E$.} {\bf (A)} The cumulative distribution of energies, $C_<(E)$ from Eq~(\ref{cume}), for the  K-pairwise  models (red) and the data (black), in a population of  120 neurons. Inset shows the high energy tails of the distribution, $C_>(E)$ from Eq~(\ref{cumge}); dashed line denotes the energy that corresponds to the probability of seeing the pattern once in an experiment.  {\bf (B)} Relative difference in the first two moments (mean, $\langle E \rangle$, dashed; standard deviation, $\sigma_E=\sqrt{\langle E^2\rangle - \langle E\rangle^2}$, solid) of the distribution of energies evaluated over real data and a sample from the corresponding model (black = pairwise; red = K-pairwise). Error bars are s.d. over 30 subnetworks at a given size $N$.\label{f8}}
\end{figure}

Maximum entropy models assign an effective energy to every possible combination of spiking and silence in the network, $E = {\mathcal H}(\{\sigma_{\rm i}\})$ from Eq~(\ref{HKpair}).    Learning the model means specifying all the parameters in this expression, so that the mapping from states to energies is completely determined.  The energy determines the probability of the state, and while we can't estimate the probabilities of all possible states, we can ask whether the distribution of energies that we see in the data agrees with the predictions of the model.  Thus, if we have a set of states drawn out of a distribution $Q(\{\sigma_{\rm i}\})$, we can count the number of states that have energies lower than $E$,
\begin{equation}
C_< (E) = \sum_{\{\sigma_{\rm i}\}}Q(\{\sigma_{\rm i}\})\Theta \left[ E-\mathcal{H}(\{\sigma_{\rm i}\})\right] , \label{cume}
\end{equation}
where $\Theta(x)$ is the Heaviside step function,
\begin{eqnarray}
\Theta (x > 0) &=& 1; \nonumber \\
\Theta (x < 0) &=& 0 .
\end{eqnarray}
Similarly, we can count the number of states that have energy larger than $E$,
\begin{equation}
C_> (E) = \sum_{\{\sigma_{\rm i}\}}Q(\{\sigma_{\rm i}\})\Theta \left[ \mathcal{H}(\{\sigma_{\rm i}\}) - E\right] , \label{cumge}
\end{equation}
Now we can take the distribution $Q(\{\sigma_{\rm i}\})$ to be the distribution of states that we actually see in the experiment, or we can take it to be the distribution predicted by the model, and if the model is accurate we should find that the cumulative distributions are similar in these two cases.  Results are shown in Fig~\ref{f8}. 

We see that the distribution of energies in the data and the model are very similar.  There is an excellent match in the ``low energy'' (high probability) region, and then as we look at the high energy tail ($C_>(E)$) we see that theory and experiment match out to probabilities of better than $C_> \sim 10^{-1}$.  Thus the distribution of energies, which is an essential construct of the model, seems to match the data across $>90\%$ of the states that we see.   

The successful prediction of the cumulative distribution $C_>(E)$ is especially striking because it extends to $E\sim 25$. At these  energies, the probability of any single state is predicted to be $e^{-25}\sim 10^{-11}$, which means that these states should occur roughly once per fifty years (!).  This seems ridiculous---what are such rare states doing in our analysis, much less as part of the claim that theory and experiment are in quantitative agreement?  The key is that there are many, many of these rare states---so many, in fact, that the theory is predicting that $\sim 10\%$ of the all the states we observe  will be (at least) this rare:  individually surprising events are, as a group, quite common.  In fact, of the $2.83\E{5}$ combinations of spiking and silence  ($1.27\pm 0.03\E{5}$ distinct ones) that we see in subnetworks of $N=120$ neurons, $1.18\pm 0.03\E{5}$ of these occur only once, which means we really don't know anything about their probability of occurrence.  We can't say that the probability of any one of these rare states is being predicted correctly by the model,  since we can't measure it, but we can say that the distribution of (log) probabilities---that is, the distribution of energies---across the set of observed states is correct, down to the $\sim 10\%$ level.    The model thus is predicting things far beyond what can be inferred directly from commonly observed patterns of activity.

\begin{figure}[b]
\centerline{\includegraphics[width=0.8\linewidth]{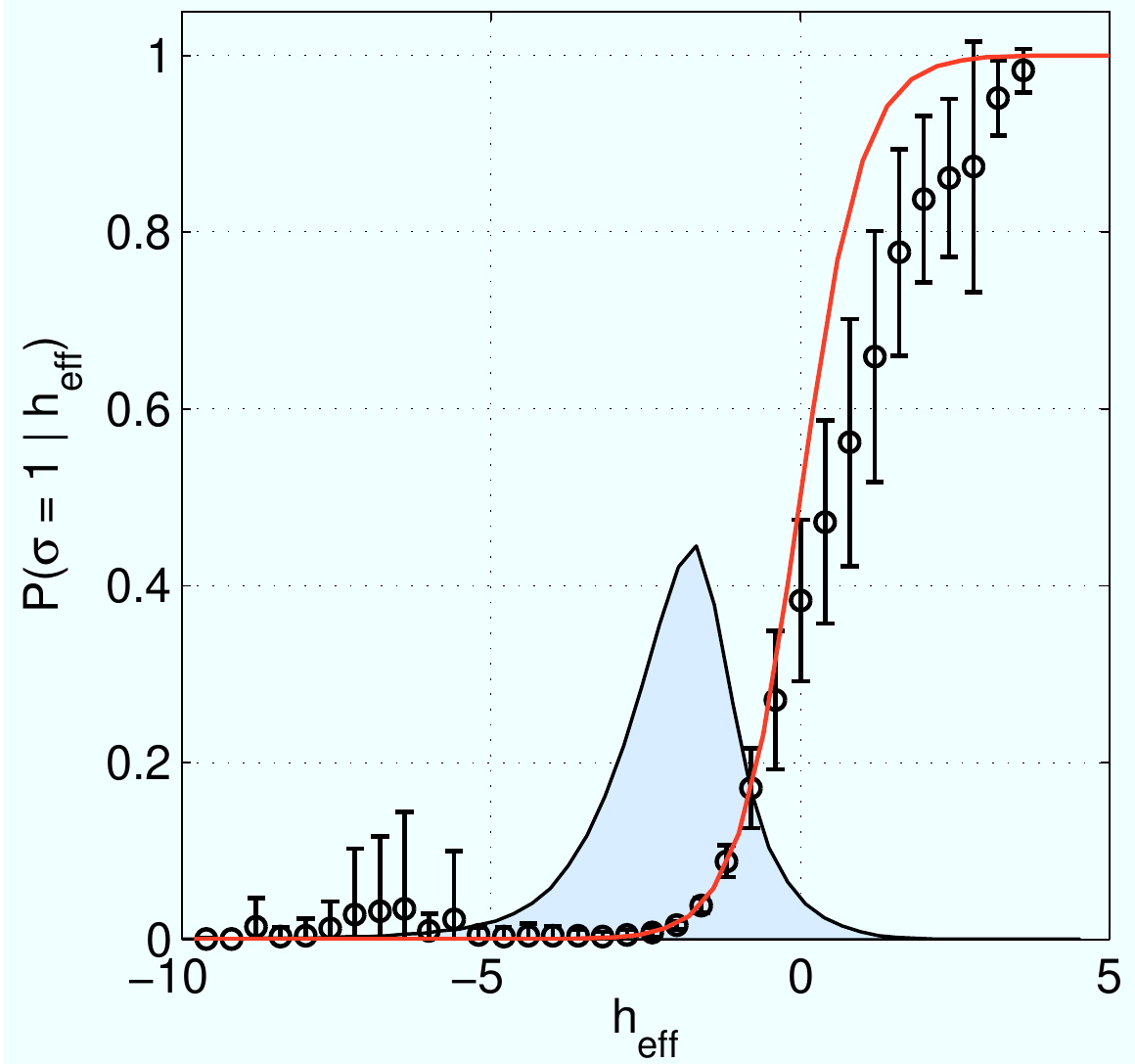}}
\caption{{\bf Effective field and spiking probabilities in a network of $N=120$ neurons.}  Given any  configuration of $N-1$ neurons, the K-pairwise model predicts the probability of firing of the $N$-th neuron by Eqs~(\ref{heff},\ref{tanh}); the effective field $h_{\rm eff}$ is fully determined by the parameters of the maximum entropy model and the state of the network. For each activity pattern in recorded data we computed the effective field, and binned these values (shown on x-axis).  For every bin we estimated from data the probability that the $N$-th neuron spiked (black circles; error bars are s.d. across 120 cells). This is compared with a parameter-free prediction (red line) from Eq~(\ref{tanh}). The gray shaded region shows the distribution of the values of $h_{\rm eff}$ over all 120 neurons and all patterns in the data.  \label{f9}}
\end{figure}

Finally, the structure of the models we are considering is that the state of each neuron---an Ising spin---experiences an ``effective field'' from all the other spins, determining the probability of spiking vs. silence.  This effective field consists of an intrinsic bias for each neuron, plus the effects of interactions with all the other neurons:
\begin{eqnarray}
h_{{\rm eff},{\rm i}} &=& \frac{1}{2}\left\{\mathcal{H}(\sigma_1,\dots,\sigma_{\rm i}=1,\dots,\sigma_{\rm N})\right. \label{heff}\\ \nonumber
&&\,\,\,\,\,\,\,\,\,\, -\left.\mathcal{H}(\sigma_1,\dots,\sigma_{\rm i}=-1,\dots,\sigma_{\rm N})\right\}.
\end{eqnarray}
  If the model is correct, then the probability of spiking is simply related to the effective field,
\begin{equation}
P(\sigma_{\rm i} = 1| h_{{\rm eff},{\rm i}} ) = {1\over{1+ e^{-h_{{\rm eff},{\rm i}} }}}.
\label{tanh}
\end{equation}
To test this relationship, we can choose one neuron, compute the effective field from the states of all the other neurons, at every moment in time,  then collect all those moments when $h_{\rm eff}$ is in some narrow range, and see how often the neuron spikes. We can then repeat this for every neuron, in turn. If the model is correct, spiking probability should depend on the effective field according to Eq~(\ref{tanh}).   We emphasize that there are no new parameters to be fit, but rather a parameter--free relationship to be tested.  The results are shown in Fig~\ref{f9}.   We see that, throughout the range of fields that are well sampled in the experiment, there is good agreement between the data and Eq~(\ref{tanh}).  As we go into the tails of the distribution, we see some deviations, but error bars also are (much) larger.  

\section{What do the models teach us?}

We have seen that it is possible to construct maximum entropy models which match the mean spike probabilities of each cell, the pairwise correlations, and the distribution of summed activity in the network, and that our data are sufficient to insure that all the parameters of these models are well determined, even when we consider groups of $N=100$ neurons or more.  Figures~\ref{f7} through \ref{f9} indicate that these models give a fairly accurate description of the distribution of states---the myriad combinations of spiking and silence---taken on by the network as a whole.  In effect we have constructed a statistical mechanics for these networks, not by analogy or metaphor but in quantitative detail.  We now have to ask what we can learn about neural function from this description.

\subsection{Basins of attraction}

In the Hopfield model, dynamics of the neural network corresponds to motion on an energy surface.  Simple learning rules can sculpt the energy surface to generate multiple local minima, or attractors, into which the system can settle.  These local minima can represent stored memories, or the solutions to various computational problems \cite{hopfield+tank_85,hopfield+tank_86}.  
If we imagine monitoring a Hopfield network over a long time, the distribution of states that it visits will be dominated by the local minima of the energy function.  Thus, even if we can't take the details of the dynamical model seriously, it  still should be true that the energy landscape determines the probability distribution over states in a Boltzmann--like fashion, with multiple energy minima translating into multiple peaks of the distribution.  

In our maximum entropy models, we find a range of $J_{\rm ij}$ values encompassing both signs (Figs~\ref{f2}D and F), as in spin glasses \cite{mezard+al_87}.  The presence of such competing interactions generates ``frustration,'' where (for example) triplets of neurons cannot find a combination of spiking and silence that simultaneously minimizes all the terms in the energy function \cite{schneidman+al_06}.  In the simplest model of spin glasses, these frustration effects, distributed throughout the system, give rise to a very complex energy landscape, with a proliferation of local minima \cite{mezard+al_87}.  Our models are not precisely Hopfield models, nor are they instances of the standard (more random) spin glass models.  Nonetheless, by looking at the pairwise $J_{\rm ij}$ terms in the energy function of our models, $48\pm 2\%$ of all interacting triplets of neurons are frustrated across different subnetworks of  various sizes ($N\geq 40$), and it is reasonable to expect that we will find many local minima in the energy function of the network.

\begin{figure}[t]
\includegraphics[width=0.7\linewidth]{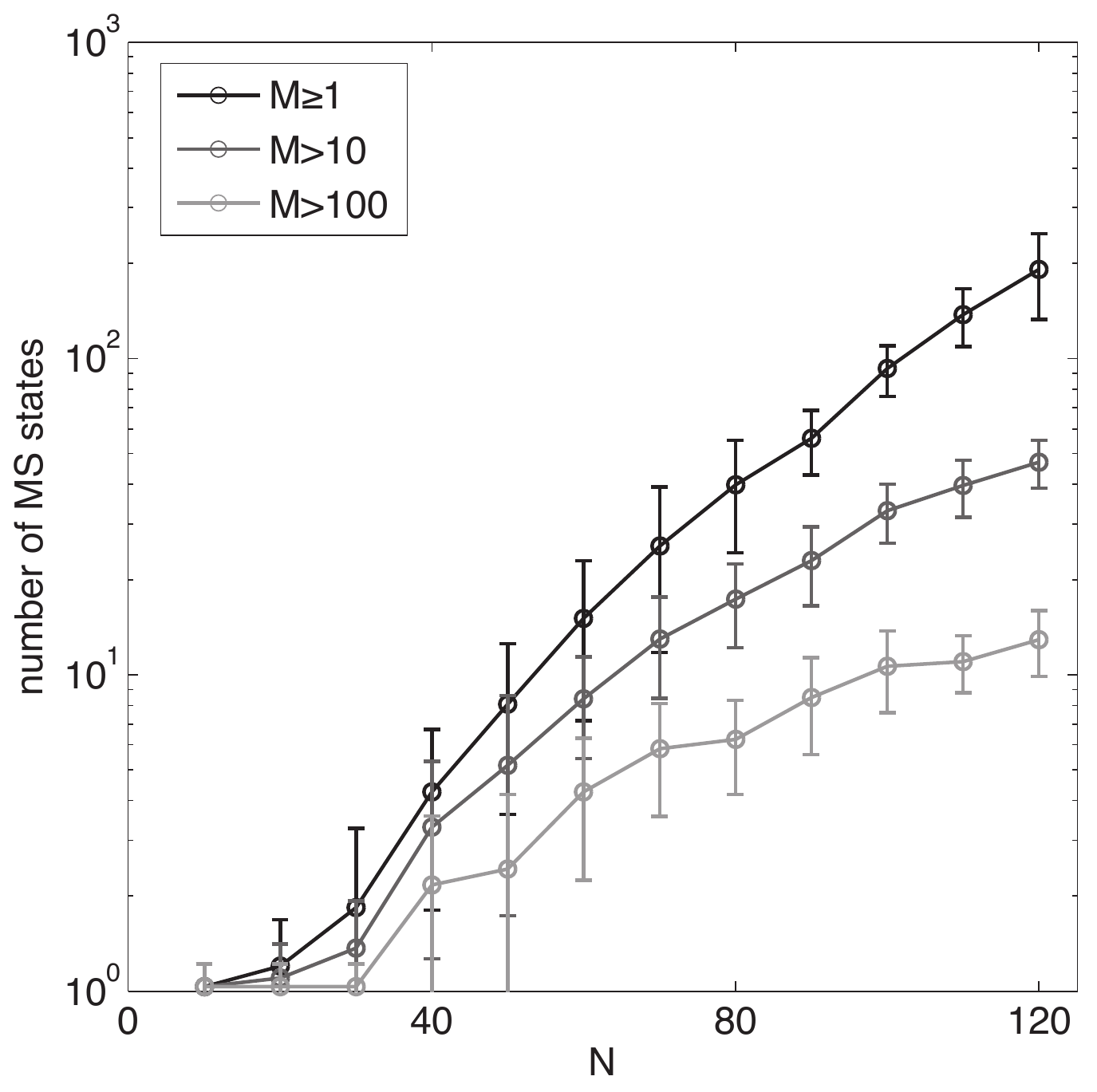} 
\caption{{\bf The number of identified metastable patterns.} Every recorded pattern is assigned to its basin of attraction by descending on the energy landscape. The number of distinct basins is shown as a function of the network size, $N$, for K-pairwise models (black line). Gray lines show the subsets of those basins that are encountered multiple times in the recording (more than 10 times, dark gray; more than 100 times, light gray). Error bars are s.d. over 30 subnetworks at every $N$. Note the logarithmic scale for the number of MS states.
}
\label{f10}
\end{figure}

\begin{figure*}[]
\includegraphics[width=7.0in]{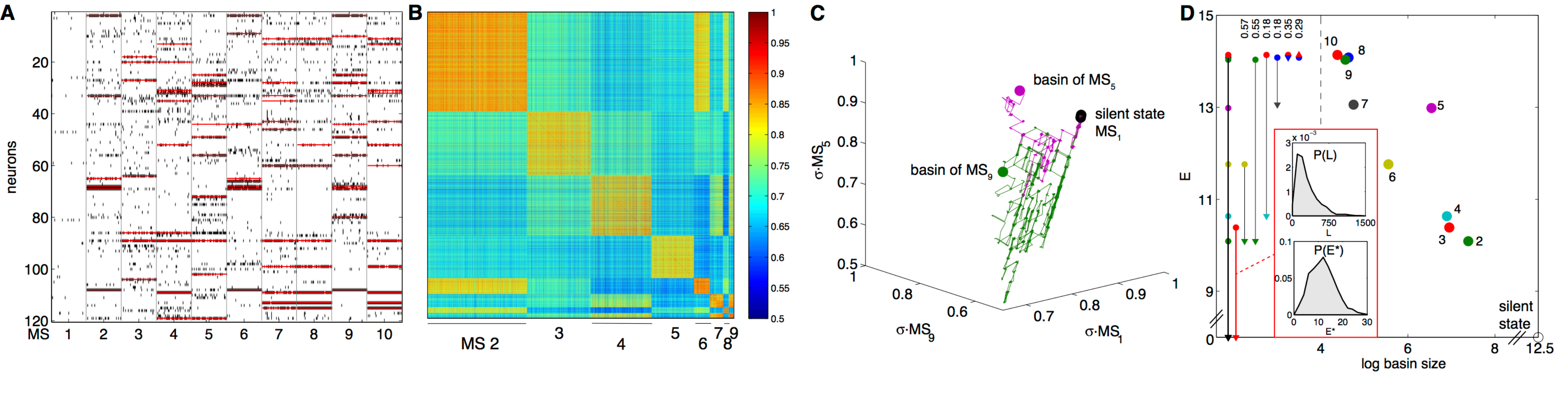} 
\caption{{\bf Energy landscape in a $N=120$ neuron K-pairwise model.}  {\bf (A)} The 10 most frequently occurring metastable (MS) states (active neurons for each in red), and 50 randomly chosen activity patterns for each MS state (black dots represent spikes). MS 1 is the all-silent basin. {\bf (B)} The overlaps, ${\cal C}_{\mu\nu}$, between all pairs of identified patterns belonging to basins $2,\dots,10$ (MS 1 left out due to its large size). Patterns within the same basin are much more similar between themselves than to patterns belonging to other basins. {\bf (C)} The structure of the energy landscape explored with Monte Carlo. Starting in the all-silent state, single spin-flip steps are taken until the configuration crosses the energy barrier into another basin. Here, two such paths are depicted (green, ultimately landing in the basin of MS 9; purple, landing in basin of MS 5) as projections into 3D space of scalar products (overlaps) with the MS 1, 5, and 9. {\bf (D)} The detailed structure of the energy landscape. 10 MS patterns from (A) are shown in the energy (y-axis) vs log basin size (x-axis) diagram (silent state at lower right corner). At left, transitions frequently observed in MC simulations starting in each of the 10 MS states, as in (C). The most frequent transitions are decays to the silent state. Other frequent transitions (and their probabilities) shown using vertical arrows between respective states. Typical transition statistics (for MS 3 decaying into the silent state) shown in the inset: the distribution of spin-flip attempts needed, $P(L)$, and the distribution of energy barriers, $P(E^*)$, over 1000 observed transitions.
}
\label{f11}
\end{figure*}

\begin{figure}[t]
\includegraphics[width=\linewidth]{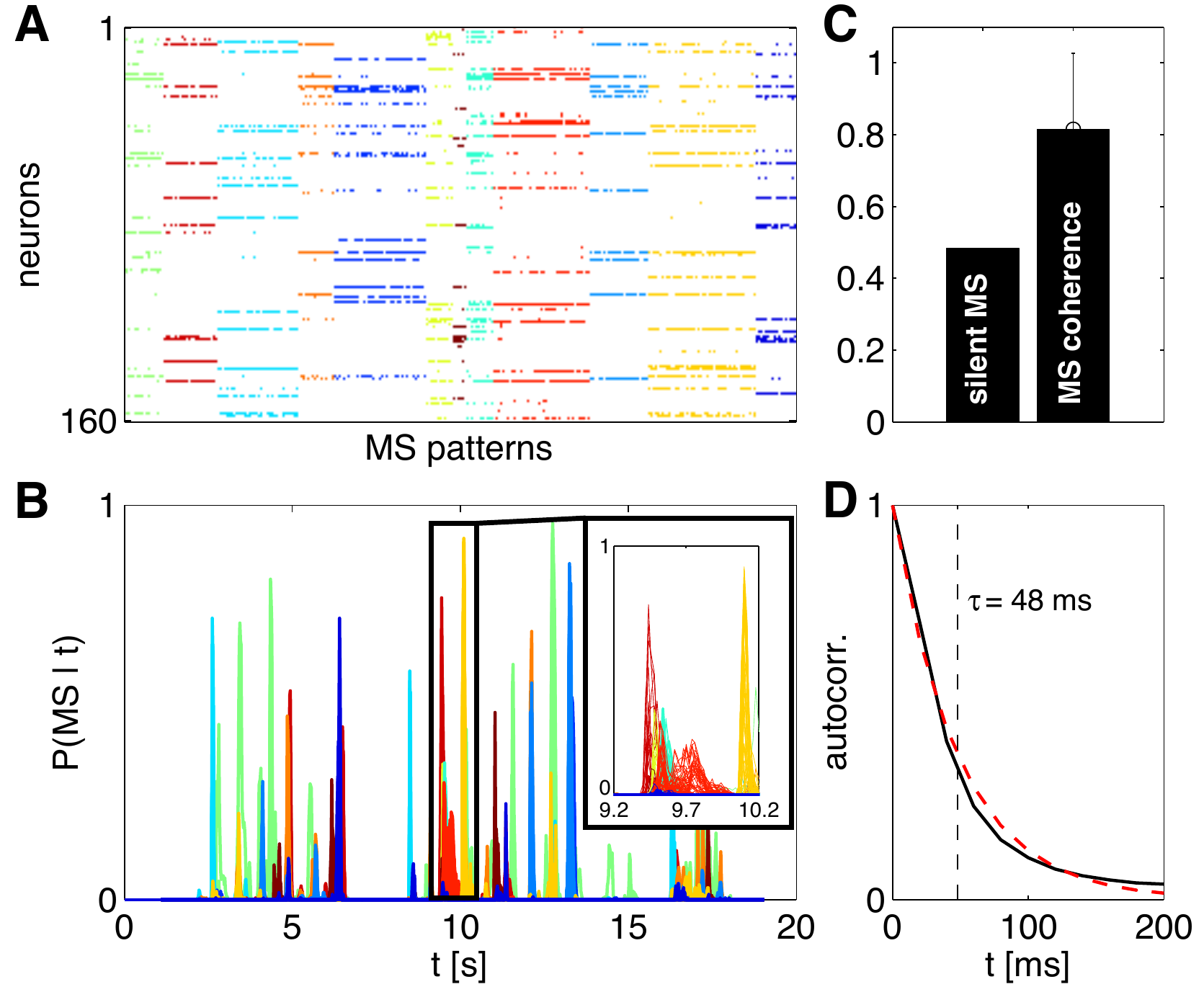} 
\caption{{\bf Basin assignments are reproducible across stimulus repeats and across subnetworks.}
{\bf (A)} Most frequently occurring MS patterns collected from 30 subnetworks of size $N=120$ out of a total population of 160 neurons; patterns have been clustered into 12 clusters (colors). {\bf (B)} The probability (across stimulus repeats) that the population is in a particular basin of attraction at any given time. Each line corresponds to one pattern from (A); patterns belonging to the same cluster are depicted in the same color. Inset shows the detailed structure of several transitions out of the all-silent state; overlapping lines of the same color show that the same transition is identified robustly across different subnetwork choices of 120 neurons out of 160. {\bf (C)} On about half of the time bins, the population is in the all-silent basin; on the remaining time bins, the coherence (the probability of being in the dominant basin divided by the probability of being in every possible non-silent basin) is high. {\bf (D)} The average autocorrelation function of traces in (B), showing the typical time the population stays within a basin (dashed red line is best exponential fit with $\tau=48\e{ms}$, or about 2.5 time bins).
}
\label{f12}
\end{figure}

To search for local minima of the energy landscape, we take every combination of spiking and silence observed in the data and move ``downhill'' on the function ${\mathcal H} (\{\sigma_{\rm i}\})$ from Eq (\ref{HKpair}) (see Appendix~\ref{app_descent}).  When we can no longer move downhill, we have identified a locally stable pattern of activity, or a ``metastable state,'' $\mathrm{MS}_\alpha=\left\{\sigma_{\rm i}^{\alpha}\right\}$, such that a flip of any single spin---switching the state of any one neuron from spiking to silent, or vice versa---increases the energy or decreases the probability of the new state.   This procedure also partitions the space of all $2^N$ possible patterns into domains, or basins of attraction, centered on the metastable states, and compresses the microscopic description of the retinal state to a number $\alpha$ identifying the basin to which that state belongs.

Figure~\ref{f10} shows how the number of  metastable states that we identify in the data grows with the size $N$ of the network. At very small $N$, the only stable configuration is the all-silent state, but for $N>30$ the metastable states start to proliferate.   Indeed, we see no sign that the number of metastable states is saturating, and the growth is certainly faster than linear in the number of neurons. Moreover, the total numbers of possible metastable states in the models' energy landscapes could be substantially higher than shown, because we only count those states that are accessible by descending  from patterns \emph{observed in the experiment}.  It thus is possible that these real networks exceed the ``capacity'' of model networks \cite{amit_89,hertz_91}.

Figure \ref{f11}A provides a more detailed view of the most prominent metastable states, and the ``energy valleys'' that surround them.  The structure of the energy valleys can be thought of as clustering the patterns of neural activity, although in contrast to the usual formulation of clustering we don't need to make an arbitrary choice of metric for similarity among patterns.  Nonetheless, we can measure the overlap ${\cal C}_{\mu \nu}$ between all pairs of  patterns $\{\sigma_{\rm i}^\mu\}$ and $\{\sigma_{\rm i}^\nu\}$ that we see in the experiment,
\begin{equation}
{\cal C}_{\mu \nu} = {1\over N} \sum_{{\rm i}=1}^N \sigma_{\rm i}^\mu \sigma_{\rm i}^\nu ,
\end{equation}
and we find that patterns which fall into the same valley are much more correlated with one another than they are with patterns that fall into other valleys (Fig \ref{f11}B).  If we start at one of the metastable states and take a random ``uphill'' walk in the energy landscape (Appendix~\ref{app_descent}), we eventually reach a transition state where there is a downhill path into other metastable states, and a selection of these trajectories is shown in Fig \ref{f11}C.  Importantly, the transition states are at energies quite high relative to the metastable states (Fig \ref{f11}D), so the peaks of the probability distribution are well resolved from one another.  In many cases it takes a large number of steps to find the transition state, so that the metastable states are substantially separated in Hamming distance.  

Individual neurons in the retina are known to generate rather reproducible responses to naturalistic stimuli \cite{puchalla+al_05,marre+al_12}, but even a small amount of noise in the response of single cells is enough to ensure that groups of $N=100$ neurons almost never generate the same response to two repetitions of the same visual stimulus.  It is striking, then, that when we show the same movie again, the retina revisits  the same basin of attraction with very high probability, as shown in Fig \ref{f12}.  The same metastable states and corresponding valleys are identifiable from different subsets of the full population, providing a measure of redundancy that we explore more fully below.  Further, the transitions into and out of these valleys are very rapid, with a time scale of just $\sim 2.5\Delta \tau$.   In summary, the neural code for visual signals seems to respect the structure inferred from the energy landscape, despite the fact that the energy landscape is constructed without reference to the visual stimuli.

\subsection{Entropy}

Central to our understanding of neural coding is the entropy of the responses \cite{spikes}. Conceptually, the entropy measures the size of the neural vocabulary:  with $N$ neurons there are $2^N$ \emph{possible} configurations of spiking and silence, but since not all of these have equal probabilities---some, like the all-silent pattern, may occur orders of magnitude more frequently than others, such as the all-spikes pattern---the \emph{effective} number of configurations is reduced to $2^{S(N)}$, where $S(N)$ is the entropy of the vocabulary for the network of $N$ neurons. Furthermore, if the patterns of spiking and silence really are codewords for the stimulus, then the mutual information between the stimulus and response, $I(\{\sigma_{\rm i}\};\mathrm{stimulus})$, can be at most the entropy of the codewords, $S[P(\{\sigma_{\rm i}\})]$.  Thus, the entropy of the system's output  bounds the information transmission. This is true even if the output words are correlated in time; temporal correlations imply that the entropy of state sequences is smaller than expected from the entropy of single snapshots, as studied here, and hence the limits on information transmission are even more stringent \cite{sdme}.

We  cannot sample the distribution---and thus estimate the entropy directly---for large sets of neurons, but we know that maximum entropy models with constraints $\{f_\mu\}$ put an upper bound to the true entropy, $S[P(\{\sigma_{\rm i}\})] \leq S[P^{(\{f_\mu\})}(\{\sigma_{\rm i}\})]$. Unfortunately, even computing the entropy of our model distribution is not simple.  Naively, we could draw samples out of the model via Monte Carlo, and since simulations can run longer than experiments, we could hope to accumulate enough samples to make a direct estimate of the entropy, perhaps using more sophisticated methods for dealing with sample size dependences \cite{nsb}.  But this is terribly inefficient (see Appendix~\ref{App:ent_comp}).  An alternative is to make more thorough use of the mathematical equivalence between maximum entropy models and statistical mechanics.

The first approach to entropy estimation involves extending  our maximum entropy models of Eq (\ref{ham}) by introducing a parameter   analogous to the temperature $T$ in statistical physics:
\begin{equation}
P^{(\{f_\mu\})}_T(\{\sigma_{\rm i}\})=\frac{1}{Z_T(\{g_\mu\})}e^{- \mathcal{H}(\{\sigma_{\rm i}\})/T}. \label{hamb}
\end{equation}
Thus, for $T=1$, the distribution in Eq (\ref{hamb}) is exactly equal  to the maximum entropy model with parameters $\{g_\mu\}$, but by varying $T$ and keeping the $\{g_\mu\}$ constant, we access a one-parameter family of distributions. Unlike in statistical physics, $T$ here is purely a mathematical device, and we do not consider the distributions at $T\neq 1$ as describing any real network of neurons. One can nevertheless compute, for each of these distributions at temperature $T$, the heat capacity $C(T)$, and then  thermodynamics teaches us that $C(T) = T \partial{S}(T)/\partial{T}$; we could thus invert this relation to compute the entropy:
\begin{equation}
S[P^{(\{f_\mu\})}]=S(T=1) =\int_0^1 \frac{C(T)}{T}\,dT. \label{sint}
\end{equation}

The heat capacity might seem irrelevant since there is no ``heat'' in our problem, but this quantity is directly related to the variance of energy, $C(T)= \sigma_E^2/T^2$, with $\sigma_E$ as in  Fig~\ref{f8}.  The energy, in turn, is related to the logarithm of the probability, and hence the heat capacity is the variance in how surprised we should be by any state drawn out of the distribution.   In practice, we can draw sample states from a Monte Carlo simulation, compute the energy of each such state, and estimate the variance over a long simulation.  Importantly, it is well known that such estimates stabilize long before we have collected enough samples to visit every state of the system \cite{mcbook}.  Thus, we start with the inferred maximum entropy model, generate a dense family of distributions at different $T$ spanning the values from 0 to 1, and, from each distribution, generate enough samples to estimate the variance of energy and thus $C(T)$; finally, we do the integral in Eq (\ref{sint}). 

Interestingly, the mapping to statistical physics gives us other, independent ways of estimating the entropy.  The most likely state of the network, in all the cases we have explored, is complete silence.  Further, in the K-pairwise models, this probability is reproduced exactly, since it is just $P_N(K=0)$.  Mathematically, this probability is given by
\begin{equation}
P_{\rm silence} = {1\over Z} \exp\left[-E({\rm silence})\right],
\end{equation}
where the energy of the silent state is easily computed from the model just by plugging in to the Hamiltonian in Eq (\ref{HKpair}); in fact we could choose our units so that the silent state has precisely zero energy, making this even easier.  But then we see that, in this model, estimating the probability of silence (which we can do directly from the data)  is the same as estimating the partition function $Z$, which usually is very difficult since it involves summing over all possible states.  Once we have $Z$, we know from statistical mechanics that
\begin{equation}
-\ln Z = \langle E \rangle - S , \label{zes}
\end{equation}
and we can estimate the average energy from a single Monte Carlo simulation of the model at the ``real'' $T=1$ (c.f. Fig~\ref{f8}).  

Finally, there are more sophisticated Monte Carlo resampling methods  that generate an estimate of the ``density of states'' \cite{wanglandau},  related to the cumulative distributions $C_<(E)$ and $C_>(E)$ discussed above, and from this density we can compute the partition function directly.  As explained in Appendix \ref{App:ent_comp}, the three different methods of entropy estimation agree to better than $1\%$ on groups of $N=120$ neurons.  

\begin{figure}[b]
\centerline{\includegraphics[width=\linewidth]{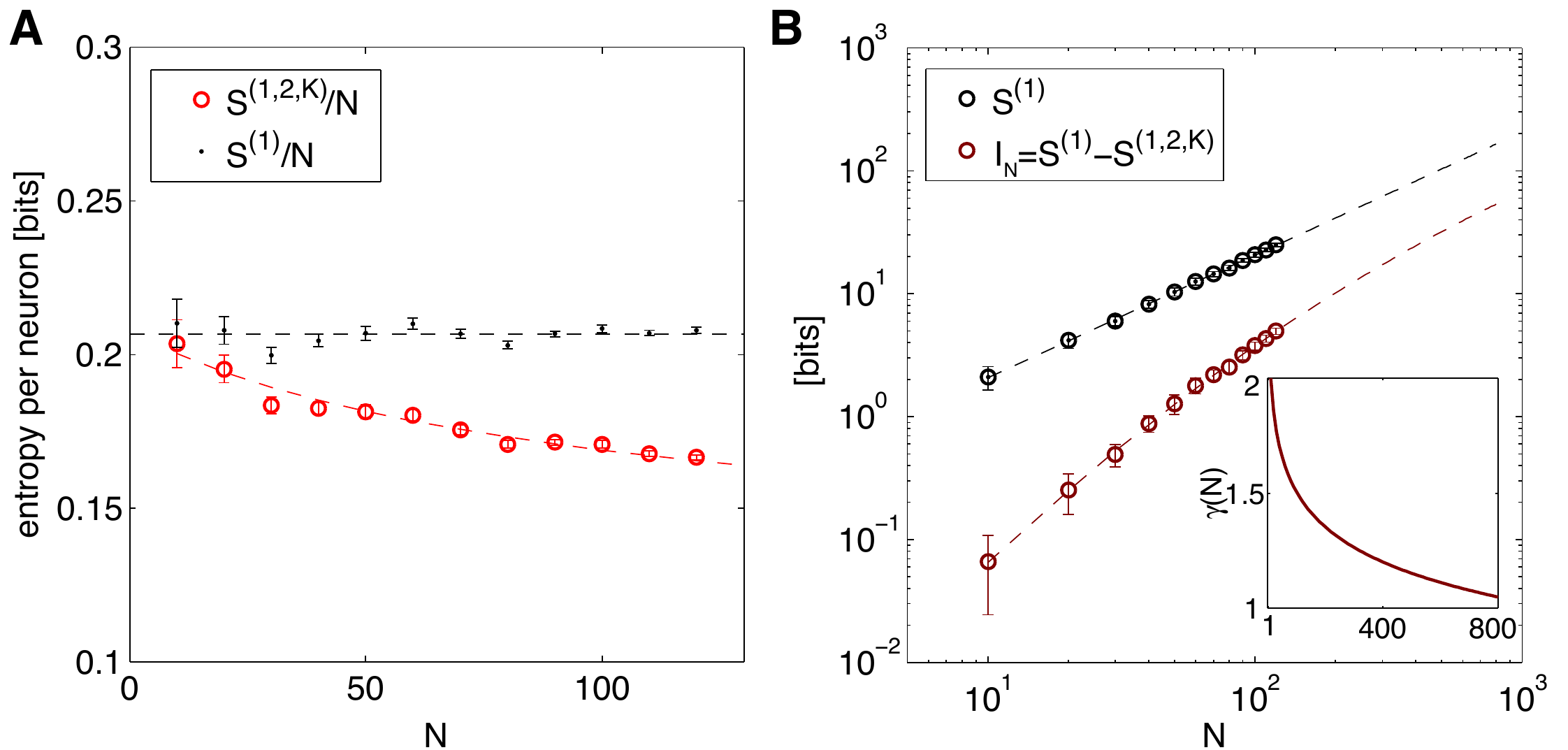} }
\caption{{\bf Entropy and multi-information from  the K-pairwise model.}  {\bf (A)} Independent entropy per neuron, $S^{(1)}/N$, in black, and the entropy of the K-pairwise models per neuron, $S^{(1,2,K)}/N$, in red, as a function of $N$. Dashed lines are fits from (B). {\bf (B)} Independent entropy scales linearly with $N$ (black dashed line). Multi-information $I_N$ of the K-pairwise models is shown in dark red. Dashed red line is a best quadratic fit for dependence of $\log I_N$ on $\log N$; this can be rewritten as $I_N\propto N^{\gamma(N)}$, where $\gamma(N)$ (shown in inset) is the effective scaling of multi-information with system size $N$. In both panels, error bars are s.d. over 30 subnetworks at each size $N$.
 \label{f13}} 
\end{figure}

Figure~\ref{f13}A shows the entropy per neuron of the K-pairwise model as a function of network size, $N$. For comparison, we also plot the independent entropy, i.e. the entropy of the non-interacting maximum entropy model that matches the mean firing rate of every neuron defined in Eq (\ref{HA}).  It is worth noting that despite the diversity of firing rates for individual neurons, and the broad distribution of correlations in pairs of neurons, the entropy per neuron varies hardly at all as we look at different groups of neurons chosen out of the larger group from which we can record.  This suggests that collective, ``thermodynamic'' properties of the network may be robust to some details of the neural population, as discussed in the Introduction.  These entropy differences between the independent and correlated models may not seem large, but losing $\Delta S = 0.05\,{\rm bits}$ of entropy per neuron means that for $N=200$ neurons the vocabulary of neural responses is restricted $2^{N\Delta S} \sim 1000$--fold.

The difference between the real entropy of the system and the independent entropy, also known as the \emph{multi--information},
\begin{equation}
I_N=S[P^{(1)}(\{\sigma_{\rm i}\})]-S[P^{(1,2,K)}(\{\sigma_{\rm i}\})],
\end{equation}
 measures the amount of statistical structure between $N$ neurons due to pairwise interactions and the K-spike constraint. As we see in Fig~\ref{f13}B,  the multi-information initially grows quadratically ($\gamma=2$) as a function of $N$. While this growth is slowing as $N$ increases, it is still faster than linear ($\gamma>1$),  and correspondingly the entropy per neuron keeps decreasing, so that even with $N=120$ neurons we have not yet reached the extensive scaling regime where the entropy per neuron would be constant. These results are consistent with suggestions in Ref \cite{schneidman+al_06} based on much smaller groups of cells; in particular the changeover towards extensive entropy growth could happen at a scale of $N\sim 200-300$, which corresponds to the total numbers of neuron within a ``correlated patch'' of this retina.  

\subsection{Coincidences and surprises}

Usually we expect that, as the number of elements $N$ in a system becomes large, the entropy $S(N)$ becomes proportional to $N$ and the distribution becomes nearly uniform over $\sim 2^{S(N)}$ states.  This is the concept of ``typicality'' in information theory \cite{cover+thomas_91} and the  ``equivalence of ensembles'' in statistical physics \cite{equiv,landau+lifshitz}.  At $N=120$, we have $S(N) = 19.97\pm 0.58$ bits, so that $2^S \sim 1\times 10^6$, and for the full $N=160$ neurons in our experiment the number of states is even larger.  In a uniform distribution, if we pick two states at random then the probability that these states are the same is given by $P_c = 2^{-S(N)}$.  On the hypothesis of uniformity, this probability is sufficiently small that large groups of neurons should {\em never} visit the same state twice during the course of a one hour experiment.  In fact, if we choose two moments in time at random from the experiment, the probability that even the full 160--neuron state that we observe will be the same is $P_c = 0.0442\pm 0.0014 $.

\begin{figure}[bt]
\includegraphics[width=\linewidth]{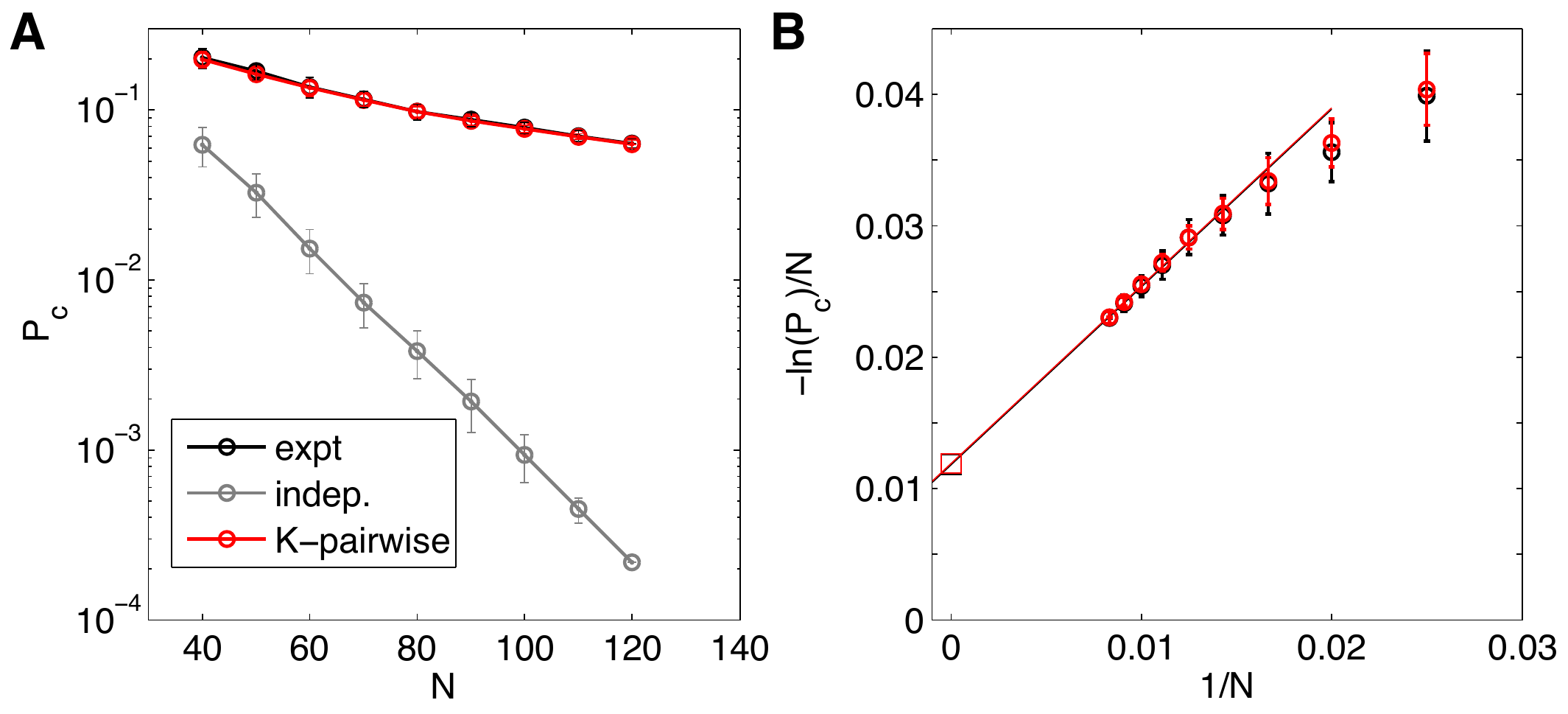} 
\caption{{\bf Coincidence probabilities.}  {\bf (A)} The probability that the combination of spikes and silences is exactly the same at two randomly chosen moments of time, as a function of the size of the population.  The real networks are orders of magnitude away from the predictions of an independent model, and this behavior is captured precisely by the K-pairwise model.  {\bf (B)}  Extrapolating the $N$ dependence of $P_c$ to large $N$.
\label{Pc}}
\end{figure}

We can make these considerations a bit more precise by exploring the dependence of coincidence probabilities on $N$.  We expect that the negative logarithm of the coincidence probability, like the entropy itself, will grow linearly with $N$; equivalently we should see an exponential decay of coincidence probability as we increase the size of the system.  This is exactly true if the neurons are independent, even if different cells have different probabilities of spiking, provided that we average over possible choices of $N$ neurons out of the population.  But the real networks are far from this prediction, as we can see in Fig \ref{Pc}A.  Larger and larger groups of neurons do seem to approach a ``thermodynamic limit'' in which $-\ln P_c \propto N$ (Fig \ref{Pc}B), but the limiting ratio $-(\ln P_c )/N = 0.0127\pm 0.0005$ is an order of magnitude smaller than our estimates of the entropy per neuron (Fig~\ref{f13}B).  Thus, the correlations among neurons make the recurrence of combinatorial patterns thousands of times more likely than would be expected from independent neurons, and this effect is even larger than simply the reduction in entropy.  This suggests that the true distribution over states is extremely inhomogeneous, not just because total silence is anomalously probable but because the dynamic range of probabilities for the different active states also is very large.  Importantly, as seen in Fig \ref{Pc}, this effect is captured with very high precision by our maximum entropy model.

\subsection{Redundancy and predictability}

\begin{figure}[b] 
\includegraphics[width=\linewidth]{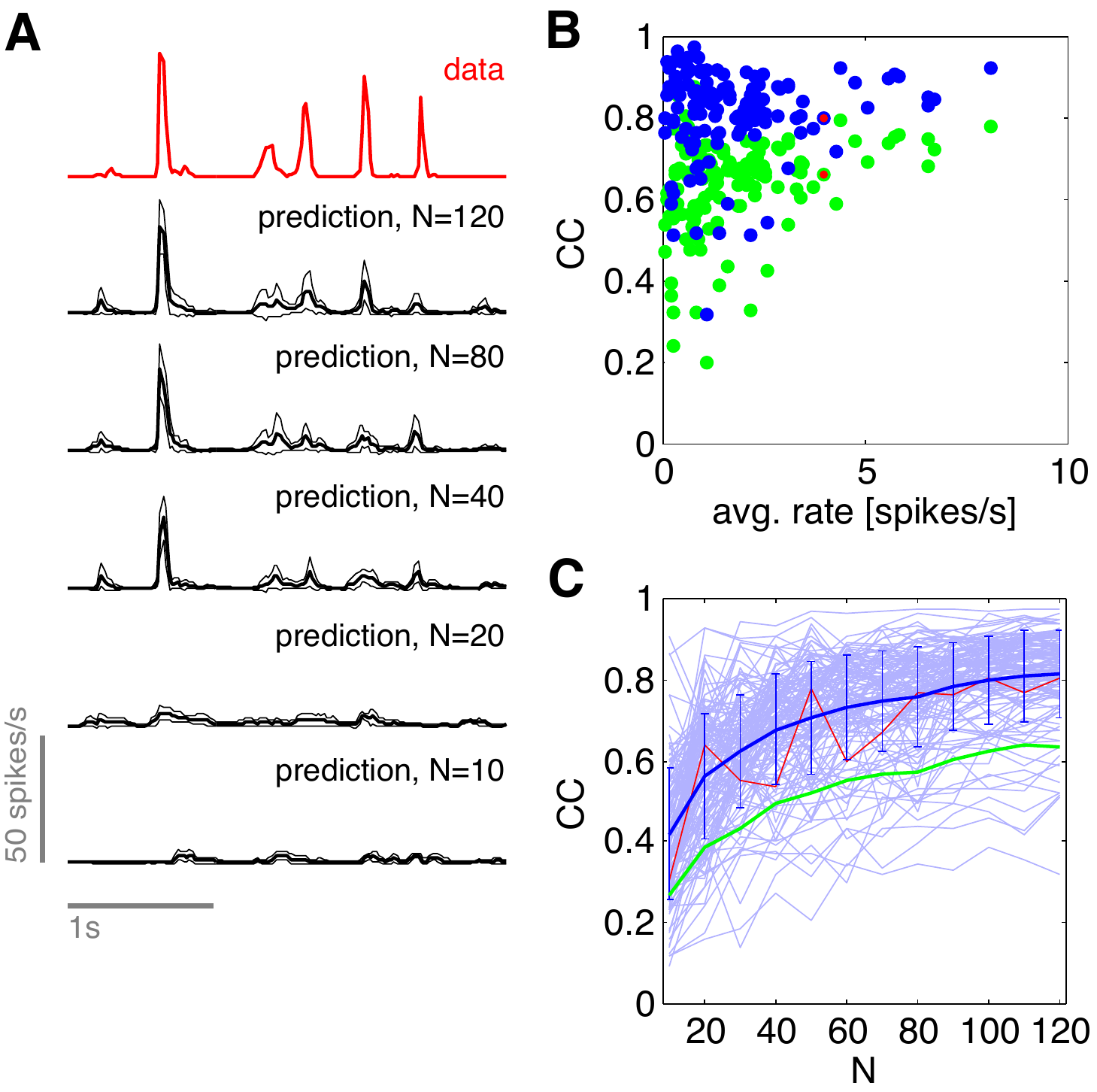} 
\caption{{\bf Predicting the firing probability of a neuron from the rest of the network.}  {\bf (A)} Probability per unit time (spike rate) of a single neuron.  Top, in red, experimental data.  Lower traces, in black, predictions based on states of other neurons in an $N$--cell group, as described in the text.  Solid lines are the mean prediction across all trials, and thin lines are the envelope $\pm$ one standard deviation.   {\bf (B)} Cross--correlation (CC) between predicted and observed spike rates vs. time, for each neuron in the $N=120$ group.  Green points are averages of CC computed from every trial, whereas blue points are the CC computed from average predictions.  {\bf (C)} Dependence of CC on the population size $N$.  Thin blue lines follow single neurons as predictions are based on increasing population sizes; red line is the cell illustrated in (A), and the line with error bars shows mean $\pm$ s.d. across all cells.  Green line shows the equivalent mean behavior computed for the green points in (B).
\label{f15}}
\end{figure}

In the retina we usually think of neurons as responding to the visual stimulus, and so it is natural to summarize their response as spike rate vs. time in a (repeated) movie, the post--stimulus time histogram (PSTH).   We can do this for each of the cells in the population that we study; one example is in the top row of Fig \ref{f15}A.  This example illustrates common features of neural responses to naturalistic sensory inputs---long epochs of near zero spike probability, interrupted by brief transients containing a small number of spikes \cite{berrypnas}.  Can our models predict this behavior, despite the fact that they make no explicit reference to the visual input?

The maximum entropy models that we have constructed predict the distribution of states taken on by the network as a whole, $P(\{\sigma_{\rm i}\})$.  From this we can construct the conditional distribution, $P(\sigma_{\rm i}|\{\sigma_{{\rm j}\neq {\rm i}}\})$, which tells us the probability of spiking in one cell given the current state of all the other cells, and hence we have a prediction for the spike probability in one neuron at each moment in time.  Further, we can repeat this construction using not all the neurons in the network, but only a group of $N$, with variable $N$. 

As the stimulus movie proceeds, all of the cells in the network are spiking, dynamically, so that the state of the system varies.   Through the conditional distribution $P(\sigma_{\rm i}|\{\sigma_{{\rm j}\neq {\rm i}}\})$, this varying state predicts a varying spike probability for the one cell in the network on which we are focusing, and we can plot this predicted probability vs. time in the same way that we would plot a conventional PSTH.  On each repeat of the movie, the states of the network are slightly different, and hence the predicted PSTH is slightly different.  What we see in Fig \ref{f15}A is that, as we use more and more neurons in the network to make the prediction, the PSTH based on collective effects alone, trial by trial, starts to look more and more like the real PSTH obtained by averaging over trials.  In particular, the predicted PSTH has near zero spike probability over most of the time, the short epochs of spiking are at the correct moments, and these epochs have the sharp onsets observed experimentally. These are features of the data which are very difficult to reproduce in models that, for example, start by linearly filtering the visual stimulus through a receptive field \cite{vanhateren,shapley+victor,victor+shapley,carandini,chen}.  In contrast, the predictions in Fig~\ref{f15} make no reference to the visual stimulus, only to the outputs of other neurons in the network.

We can evaluate the predictions of spike probability vs. time by computing the correlation coefficient between our predicted PSTH and the experimental PSTH, as has been done in many other contexts \cite{vanhateren,dan+reid,th1}.  Since we generate a prediction for the PSTH on every presentation of the movie, we can compute the correlation from these raw predictions, and then average, or average the predictions and then compute the correlation; results are shown in Figs \ref{f15}B and C.  We see that correlation coefficients can reach $\sim 0.8$, on average, or even higher for particular cells.  Predictions seem of more variable quality  for cells with lower average spike rate, but this is a small effect. The quality of average predictions, as well as the quality of single trial predictions,  still seem to grow gradually as we include more neurons even at $N\sim 100$, so it may be that we have not seen the best possible performance yet.

Our ability to predict the state of individual neurons by reference to the network, but not the visual input, means that the representation of the sensory input in this population is substantially redundant.   Stated more positively, the full information carried by this population of neurons---indeed, the full information available to the brain about this small patch of the visual world---is accessible to downstream cells and areas that receive inputs from only a fraction of the neurons.

\section{Discussion}

It is widely agreed that  neural activity in the brain is more than the sum of its parts---coherent percepts, thoughts, and actions require the coordinated activity of many neurons in a  network, not the independent activity of many individual neurons.    It is not so clear, however, how to build bridges between this intuition about collective behavior and the activity of individual neurons.

One set of ideas is that the activity of the network as a whole may be confined to  some very low dimensional trajectory, such as a global, coherent oscillation.  Such oscillatory activity is observable in the summed electrical activity of large numbers of neurons---the EEG---and should be reflected as oscillations in the (auto--)correlation functions of spike trains from individual neurons. On a more refined level, dimensionality reduction techniques like PCA allow the activity patterns of a neural network to be viewed on a low-dimensional manifold, facilitating visualization and intuition \cite{stopfer+laurent_03,brody+al_2010, churchland+shenoy_10,laurent+al_13}. A very different idea is provided by the Hopfield model, in which collective behavior is expressed in the stabilization of many discrete patterns of activity, combinations of spiking and silence across the entire network \cite{hopfield_82,amit_89}.  Taken together, these many patterns can span a large fraction of the full space of possibilities, so that there need be no dramatic dimensionality reduction in the usual sense of this term.  

The claim that a network of neurons exhibits collective behavior is really the claim that the distribution of states taken on by the network has some nontrivial structure that cannot be factorized into contributions from individual cells or perhaps even smaller subnetworks.  Our goal in this work has been to build a model of this distribution, and to explore the structure of that model.  We emphasize that building a model is, in this view, the first step rather than the last step.  But building a model is challenging, because the space of states is very large and data are limited.

An essential step in searching for collective behavior has been to develop experimental techniques that allow us to record not just from a large number of neurons, but from a large fraction of the neurons in a densely interconnected region of the retina \cite{segev+al_04,marre+al_12}.    In large networks, even measuring the correlations among pairs of neurons can become problematic:  individual elements of the correlation matrix might be well determined from small data sets, but much larger data sets are required to be confident that the matrix as a whole is well determined.  Thus, long, stable recordings are even more crucial than usual.    

To use the maximum entropy approach, we have to be sure that we can actually find the models that reproduce the observed expectation values (Fig~\ref{f2},~\ref{f3}) and that we have not, in the process, fit to spurious correlations that arise from the finite size of our data set (Fig~\ref{f4}).  Once these tests are passed, we can start to assess the accuracy of the model as a description of the network as a whole. In particular, we found that the pairwise model began to break down at a network size $N\geq 40$ (Fig~\ref{f5}). However, by adding the constraint that  reproduces the probability of $K$ out of $N$ neurons spiking synchronously (Fig~\ref{f6}), which is a statistic that is well-sampled and does not greatly increase the model's complexity, we could again recover good performance (Figs~\ref{f7}-\ref{f9}).

Perhaps the most useful global test of our models is to ask about the distribution of state probabilities:  how often should we see combinations of spiking and silence that occur with probability $P$?  This has the same flavor as asking for the probability of every state, but  does not suffer from the curse of dimensionality.  Since maximum entropy models are mathematically identical to the Boltzmann distribution in statistical mechanics, this question about the frequency of states with probability $P$ is the same as asking how many states have a given energy $E$; we can avoid binning along the $E$ axis by asking for the number of models with energies smaller (higher probability) or larger (lower probability) than $E$.  Figure \ref{f8} shows that these cumulative distributions computed from the model agree with experiment far into the tail of low probability states.  These states are so rare that, individually, they almost never occur, but there are so many of these rare states that, in aggregate, they make a measurable contribution to the distribution of energies.  Indeed, most of the states that we see in the data are rare in this sense, and their statistical weight is correctly predicted by the model.  

The maximum entropy models that we construct from the data do not appear to simplify along any conventional axes.  The matrix of correlations among spikes in different cells (Fig \ref{f1}A) is of full rank, so that principal component analysis does not yield significant dimensionality reduction.  The matrix of ``interactions'' in the model (Fig \ref{f1}D) is neither very sparse nor of low rank, perhaps because we are considering a group of neurons all located (approximately) within the radius of the typical dendritic arbor, so that all cells have a chance to interact with one another.    Most importantly, the interactions that we find are not weak (Fig \ref{f1}F), and together with being widespread this means that their impact is strong.  Technically, we cannot capture the impact within low orders of perturbation theory (Appendix \ref{App:perturb}), but qualitatively this means that the behavior of the network as a whole is not in any sense ``close'' to the behavior of non--interacting neurons.  Thus,  if our models work, it is not simply because correlations are weak, as had been suggested \cite{roudi2}.

Having convinced ourselves that we can build models which give an accurate description of the probability distribution over the states of spiking and silence in the network, we can ask what these models teach us about  function.  As emphasized in Ref \cite{schneidman+al_06}, one corollary of collective behavior is the possibility of error correction or pattern completion---we can predict the spiking or silence of one neuron by knowing the activity of all the other neurons.  With a population of $N=100$ cells, the quality of these predictions becomes quite high  (Fig \ref{f15}).  The natural way of testing these predictions is to look at the probability of spiking vs. time in the stimulus movie.  Although we make no reference to the stimulus, we reproduce  the sharp peaks of activity and extended silences that are so characteristic of the response to naturalistic inputs, and so difficult to capture in conventional models where each individual neuron responds to the visual stimulus as seen through its receptive field \cite{vanhateren}.  

One of the dominant concepts in thinking about the retina has been the idea that the structure of receptive fields serves to reduce the redundancy of natural images and enhance the efficiency of information transmission to the brain \cite{barlow,attneave,atick_redlich,vanhateren1} (but see \cite{puchalla+al_05,barlow_revisited}).  While one could argue that the observed redundancy among neurons is less than expected from the structure of natural images or movies, none of what we have described here would happen if the retina truly ``decorrelated'' its inputs.  Far from being almost independent, the combinations of spiking and silence in different cells cluster into basins of attraction defined by the local minima of energy in our models, and the spiking of each neuron is highly predictable from the activity of other neurons in the network, even without reference to the visual stimulus (Fig \ref{f15}).

With $N=120$ neurons,  our best estimate of the entropy  corresponds to significant occupancy of roughly one million distinct combinations of spiking and silence.  Each state could occur with a different probability, and (aside from normalization) there are no constraints---each of these probabilities could be seen as a separate parameter describing the network activity.  It is appealing to think that there must be some simplification, that we won't need a million parameters, but it is not obvious that any particular simplification strategy will work.  Indeed, there has been the explicit claim that the approach we have taken here will not work for large networks \cite{roudi2}.  Thus, it may seem surprising that we can write down a relatively simple model, with all parameters determined from experiment, and have this model predict so much of the structure in the distribution of states.  Surprising or not, it certainly is important that, as the community contemplates monitoring the activity of ever larger number of neurons \cite{bam}, we can identify  theoretical approaches that have the potential to tame the complexity of these large systems.

Some cautionary remarks about the interpretation of our models seem in order.  Using the maximum entropy method  does not mean there is some hidden force maximizing the entropy of neural activity, or that we are describing neural activity as being in something like thermal equilibrium; all we are doing is building maximally agnostic models of the probability distribution over states.  Even in the context of statistical mechanics, there are infinitely many models for the dynamics of the system that will be consistent with the equilibrium distribution, so we should not take the success of our models to mean that the dynamics of the network corresponds to something like Monte Carlo dynamics on the energy landscape.  It is tempting to look at the   couplings $J_{\rm ij}$ between different neurons as reflecting genuine, mechanistic interactions, but even in the context of statistical physics we know that this interpretation need not be so precise:  we can achieve a very accurate description of the collective behavior in large systems even if we do not capture every microscopic detail, and the interactions that we do describe in the most successful of models often are effective interactions mediated by degrees of freedom that we need not treat explicitly.  Finally, the fact that a maximum entropy model which matches a particular set of experimental observations is successful does not mean that this choice of observables (e.g., pairwise correlations) is unique or minimal.  For all these reasons, we do not think about our models in terms of their parameters, but rather as a description of the probability distribution $P(\{\sigma_{\rm i}\})$ itself, which encodes the collective behavior of the system.

The striking feature of the distribution over states is its extreme inhomogeneity.  The entropy of the distribution is not that much smaller than it would be if the neurons made independent decisions to spike or be silent, but the shape of the distribution is very different; the network builds considerable structure into the space of states, without sacrificing much capacity.  The probability of the same state repeating is many orders of magnitude larger than expected for independent neurons, and this really is quite startling (Fig \ref{Pc}).  If we extrapolate to the full population of $\sim 250$ neurons in this correlated, interconnected patch of the retina, the probability that two randomly chosen states of the system are the same is roughly one percent.  Thus, some combination of spiking and silence across this huge population should repeat exactly every few seconds.  This is true despite the fact that we are looking at the entire visual representation of a small patch of the world, and the visual stimuli are fully naturalistic.  Although complete silence repeats more frequently,  a wide range of other states also recur, so that many different combinations of spikes and silence occur often enough that we (or the brain) can simply count them to estimate their probability.  This would be absolutely impossible in a population of nearly independent neurons,  and it has been suggested that these repeated patterns provide an anchor for learning \cite{ganmor+al_11}.  It is also possible that the detailed structure of the distribution, including its inhomogeneity, is matched to the statistical structure of visual inputs in a way that goes beyond the idea of redundancy reduction, occupying a regime in which strongly correlated activity is an optimal code \cite{tkacik+al_10,stephens+al_13,saremi+al_13}.  

Building a precise model of activity patterns required us to match the statistics of global activity (the probability that $K$ out of $N$ neurons spike in the same small window of time).  Elsewhere we have explored a very simple model in which we ignore the identity of the neurons and match only this global behavior \cite{simplest}.  This model already has a lot of structure, including the extreme inhomogeneity that we have emphasized here.  In the simpler model we can exploit the equivalence between maximum entropy models and statistical mechanics to argue that this inhomogeneity is equivalent to the statement that the population of neurons is poised near  a critical surface in its parameter space, and we have seen hints of this from analyses of smaller populations as well \cite{tkacik+al_06,tkacik+al_09}.  The idea that biological networks might organize themselves to critical points has a long history, and several different notions of criticality have been suggested \cite{mora+bialek_11}.   A sharp question, then, is whether the full probability distributions that we have described here correspond to a critical system in the sense of statistical physics, and whether we can find more direct evidence for criticality in the data, perhaps without the models as intermediaries.   

Finally, we note that the our approach to building models for the activity of the retinal ganglion cell population is entirely unsupervised:  we are making use only of structure in the spike trains themselves, with no reference to the visual stimulus.  In this sense, the structures that we discover here are structures that could be discovered by the brain, which has no access to the visual stimulus beyond that provided by these neurons.  While there are more structures that we could use---notably, the correlations across time---we find it remarkable that so much is learnable from just an afternoon's worth of data.  As it becomes more routine to record the activity of such (nearly) complete sensory representations, it will be interesting to take the organism's point of view \cite{spikes} more fully, and try to extract meaning from the spike trains in an unsupervised fashion.

\begin{acknowledgments}
We thank A Cavagna, I Giardina, T Mora, SE Palmer, GJ Stephens, and A Walczak for many helpful discussions.  This work was supported in part by  NSF Grants IIS--0613435, PHY--0957573, and CCF--0939370, and by NIH  Grants R01 EY14196 and P50 GM071508. Additional support was provided by the Fannie and John Hertz Foundation, the Swartz Foundation, and the WM Keck Foundation.  
\end{acknowledgments}

\appendix

\setcounter{figure}{0}
\makeatletter 
\renewcommand{\thefigure}{S\@arabic\c@figure} 

\section{Experimental methods}

{\bf Electrophysiology.} We analyzed the recordings from the tiger salamander (\emph{Ambystoma tigrinum}) retinal ganglion cells responding to naturalistic movie clips, as in the experiments of  Ref.~\cite{schneidman+al_06,puchalla+al_05,marre+al_12}. In brief, animals were euthanized according to institutional animal care standards. The retina was isolated from the eye under dim illumination and transferred as quickly as possible into oxygenated Ringer's medium, in order to optimize the long-term stability of recordings. Tissue was flatted and attached to a dialysis membrane using polylysine. The retina was then lowered with the ganglion cell side against a multi-electrode array. Arrays were first fabricated in university cleanroom facilities \cite{dario}.  Subsequently, production was contracted out to a commercial MEMS foundry for higher volume production (Innovative Micro Technologies, Santa Barbara, CA). Raw voltage traces were digitized and stored for off-line analysis using a 252-channel preamplifier (MultiChannel Systems, Germany). The recordings were sorted using custom spike sorting software developed specifically for the new dense  array \cite{marre+al_12}. 234 neurons passed the standard tests for the waveform stability and the lack of refractory period violations. Of those, 160 cells whose firing rates were most stable across stimulus repeats were selected for further analysis. Within this group, the mean fraction of interspike intervals (ISI) shorter than $2\e{ms}$ (i.e., possible refractory violations) was $1.3\E{-3}$.

{\bf Stimulus display.} The stimulus consisted of  a short ($t=19\e{s}$) grayscale movie clip of swimming fish and water plants in a fish tank, which was repeated 297 times. The stimulus was presented using standard optics, at a rate of 30 frames per second, and gamma corrected for the display.

{\bf Data preparation.} We randomly selected 30 subgroups of $N=10,\, 20,\, \cdots,\, 120$ cells for analysis from the total of 160 sorted cells. In sum, we  analyzed $30\times 12 = 360$ groups of neurons, which we denote by $\mathcal{S}_N^\nu$, where $N$ denotes the subgroup size, and $\nu=1,\,\cdots,\,30$ indexes the chosen subgroup of that size. Time was discretized into $\Delta t =20\e{ms}$ time bins, as in our previous work \cite{schneidman+al_06,tkacik+al_06,tkacik+al_09}. The state of the retina was represented by $\sigma_{\rm i}(t)=+1 (-1)$ if the neuron i spiked at least once (was silent) in a given time bin $t$.  This binary description is incomplete only in $\sim 0.5\%$ of the time bins that contain more than one spike; we treat these bins as $\sigma_{\rm i} = +1$. Across the entire experiment, the mean probability of non-silence (that is, $\sigma_{\rm i} = +1$) is  $\sim 3.1\%$. Time discretization resulted in 953 time bins per stimulus repeat;  $297$ presented repeats yielded a total of $T=283,041$ $N$-bit binary  samples during the course of the experiment for each subgroup.

\section{Learning maximum entropy models from data}
\label{App_EntEst}
We used a modified version of our previously published learning procedure to compute the maximum entropy models given measured constraints \cite{Broderick+al_08}; the proof of convergence for the core of this L1-regularized maximum entropy algorithm is given in Ref.~\cite{dudik}. Our new algorithm can use as constraints arbitrary functions, not only single and pairwise marginals as before. Parameters of the Hamiltonian are learned  sequentially in an order which greedily optimizes a bound on the log likelihood, and we use a variant of histogram Monte Carlo to estimate the  values of constrained statistics during learning steps \cite{swendsen}. Monte Carlo induces sampling errors on our estimates of these statistics, which provide an implicit regularization for the parameters of the Hamiltonian \cite{dudik}. We verified the correctness of the algorithm explicitly for groups of 10 and 20 neurons where exact numerical solutions are feasible. We also verified that our MC sampling had a long enough ``burn-in'' time to equilibrate, even for groups of maximal size ($N=120$), by starting the sampling repeatedly from same vs random different initial conditions (100 runs each) and  comparing the constrained statistics, as well as the average and variance of the energy and magnetization, across these runs; all statistics were not significantly dependent on the initial state (two--sample Kolmogorov-Smirnov test at significance level 0.05).

\begin{figure}[bt]
\includegraphics[width=\linewidth]{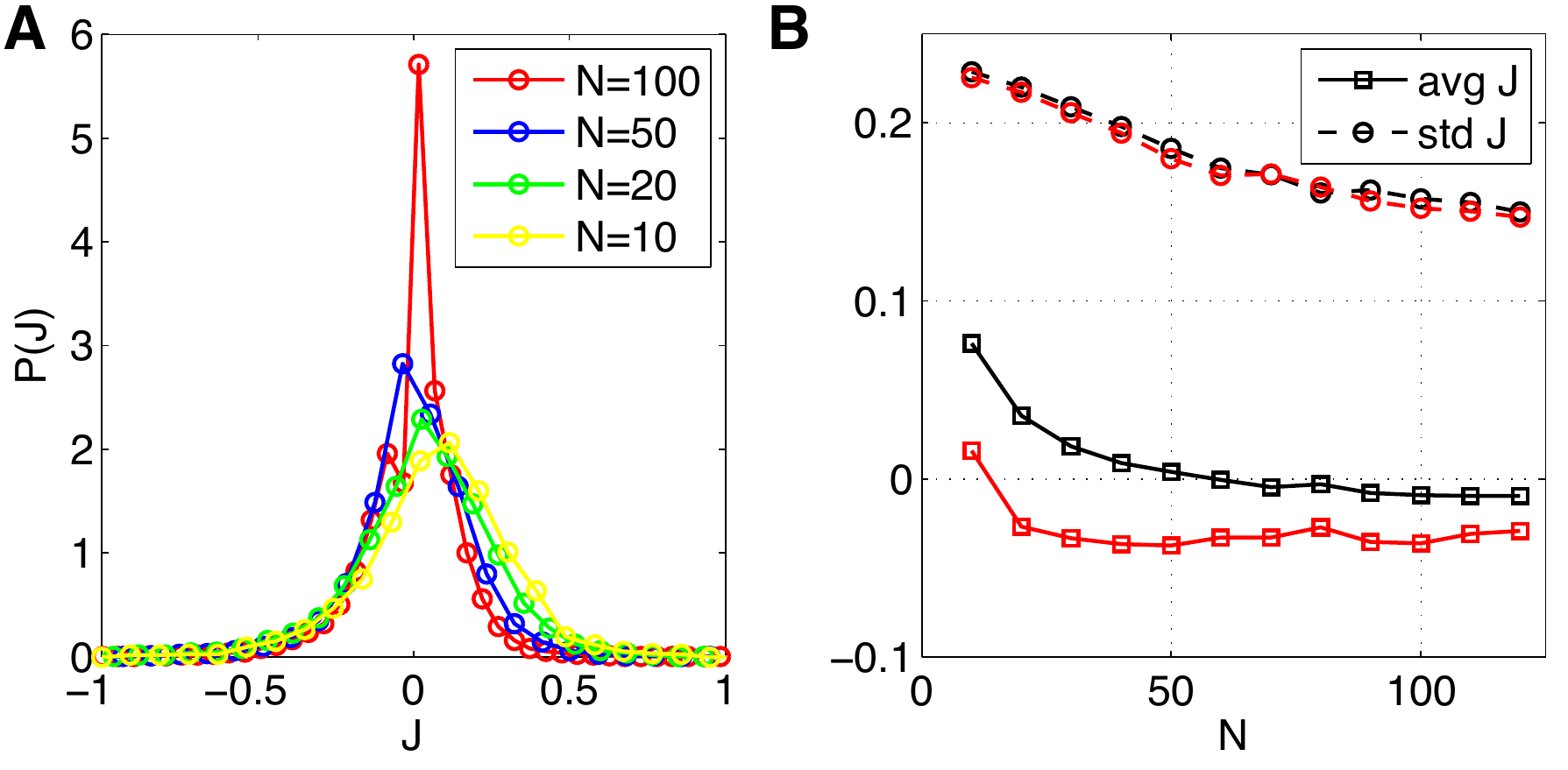} 
\caption{{\bf Interactions in the (K-)pairwise model.} {\bf (A)} The distributions of pairwise couplings, $J_{\rm ij}$, in pairwise models of Eq~(\ref{Hpair}), for different network sizes ($N$). The distribution is pooled over 30 networks at each $N$. {\bf (B)} The mean (solid) and s.d. (dashed) of the distributions in (A) as a function of network size (black); the mean and s.d. of the corresponding distributions for K-pairwise models as a function of network size (red).}
\label{sf1}
\end{figure}

Figure \ref{sf1} provides a summary of the models we have learned for populations of different sizes.  In small networks there is a systematic bias to the distribution of $J_{\rm ij}$ parameters, but as we look to larger networks this vanishes and the distribution of $J_{\rm ij}$ becomes symmetric.  Importantly, the distribution remains quite broad, with the standard deviation of $J_{\rm ij}$ across all pairs declining only slightly.  In particular, the typical coupling does not decline as $\sim 1/\sqrt{N}$, as would be expected in conventional spin glass models \cite{mezard+al_87}.  This implies, as emphasized previously \cite{tkacik+al_09}, that the ``thermodynamic limit'' (very large $N$) for these systems will be different from what we might expect based on traditional physics examples.

We withheld a random selection of 20 stimulus repeats (test set) for model validation, while training the model on the remaining 277 repeats. On training data, we computed the constrained statistics (mean firing rates, covariances, and the k-spike distribution), and used bootstrapping to estimate the error bars on each of these quantities; the constraints were the only input to the learning algorithm. Figure~\ref{f1} shows an example reconstruction for a pairwise model for $N=100$ neurons; the precision of the learning algorithm is shown in Fig.~\ref{f2}.

The dataset consists of a total of $T\sim 300\E{3}$ binary pattern samples, but due to the structure of the stimulus, the number of statistically independent samples must be smaller: while the repeats are statistically independent, the samples within each repeat are not, because they are driven by the stimulus. The  variance for a binary variable given its mean, $\langle \sigma_{\rm i}\rangle$,  is $\sigma_{\rm i,1}^2=1-\langle \sigma_{\rm i}\rangle^2$; with $R$ independent repeats, the error on the estimate in the average rate should decrease as $\sigma_{\rm i, R}^2=\sigma_{\rm i,1}^2/R$. By repeatedly estimating the statistical errors with different subsets of repeats and comparing the expected scaling of the error in the original data set and in the data set where we shuffle time bins randomly, thereby destroying the repeat structure, we can estimate the effective number of independent samples; we find this to be $T_{\rm indep}\sim 110\E{3}$, about $37\%$ of the total number of samples, $T$.

We note that our largest models have $<8\E{3}$ constrained statistics that are estimated from at least $15\times$ as many statistically independent samples. Moreover, the vast majority of these statistics are pairwise correlation coefficients that can be estimated extremely tightly from the data, often with relative errors below $1\%$, so we do not expect overfitting on general grounds. Nevertheless, we explicitly checked that there is no overfitting by comparing the log likelihood of the data under the learned maximum entropy model, for each of the 360 subgroups $\mathcal{S}_N^\nu$, on the training and testing set, as shown in Fig.~\ref{f3}. 

\section{Exploring the energy landscape}
\label{app_descent}
To find the metastable (MS) states, we start with a pattern $\{\sigma_{\rm i}\}$ that appears in the data, and attempt to flip spins ${\rm i}=1,\,\cdots,\,N$ from their current state into $-\sigma_{\rm i}$, in order of increasing $\rm i$. A flip is retained if the energy of the new configuration is smaller than before the flip. When none of the spins can be flipped, the resulting pattern is recorded as the MS state. The set of MS states found can depend on the manner in which descent is performed, in particular when some of the states visited during descent are on the ``ridges'' between multiple basins of attraction. Note that whether a pattern is a MS state or not is independent of the descent method; what depends on the method is which MS states are found by starting from the data patterns. To explore the structure of the energy landscape in Fig~\ref{f11}, we started 1000 Metropolis MC simulations repeatedly in each of the 10 most common MS states of the model; after each attempted spin-flip, we checked whether the resulting state is still in the basin of attraction of the starting MS state (by invoking the descent method above), or whether it has crossed the energy barrier into another basin. We histogrammed the transition probabilities into other MS basins of attraction and, for particular transitions, we tracked the transition paths to extract the number of spin-flip attempts and the energy barriers. ``Basin size'' of a given MS state is the number of patterns \emph{in the recorded data} from which the given MS state is reached by descending on the energy landscape. The results presented in Fig~\ref{f11} are typical of the transitions we observe across multiple subnetworks of 120 neurons.

\section{Computing the entropy and partition function of the maximum entropy distributions}
\label{App:ent_comp}
\begin{figure}[t]
\includegraphics[width=\linewidth]{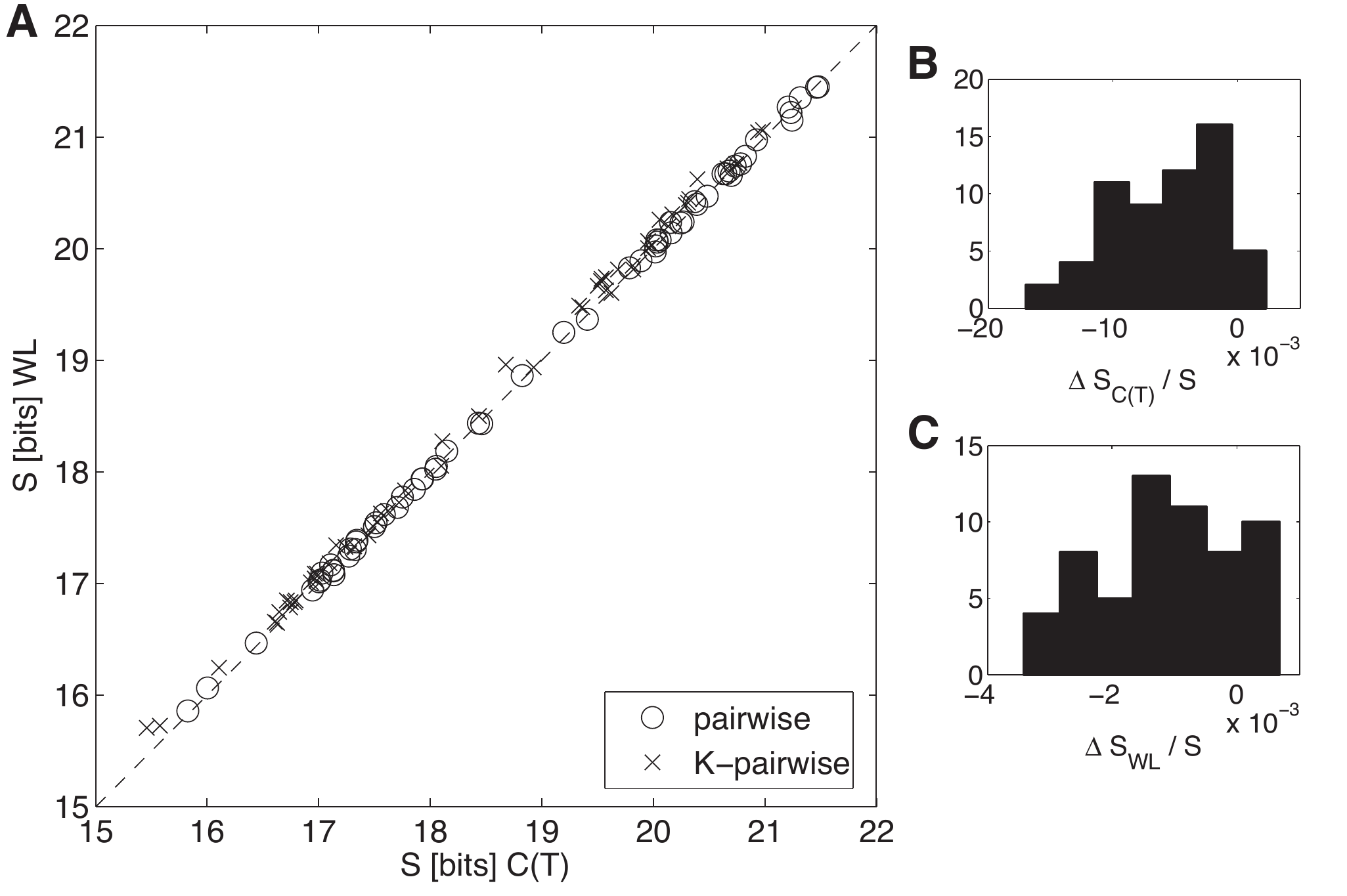} 
\caption{{\bf Precision of entropy estimates.} {\bf (A)} Entropy estimation using heat capacity integration (x-axis) from Eq~(\ref{sint}) versus entropy estimation using the Wang-Landau sampling method (y-axis) \cite{wanglandau}. Each plot symbol is one subnetwork of either $N=100$ or $N=120$ neurons (circles = pairwise models, crosses = K-pairwise models). The two sampling methods yield results that agree to within $\sim 1\%$. {\bf (B)} Fractional difference between the heat capacity method and the entropy determined from the all-silent pattern. The histogram is over 30 networks at $N=100$ and 30 at $N=120$, for the K-pairwise model. {\bf (C)} Fractional difference between the Wang-Landau sampling method and the entropy determined from the all-silent pattern. Same convention as in (B).}
\label{sf3}
\end{figure}

Entropy estimation is a challenging problem.  As explained in the text, the usual approach of counting samples and identifying frequencies with probabilities will fail catastrophically in all the cases of interest here, even if we are free to draw samples from out model rather than from real data.  Within the framework of maximum entropy models, however, the equivalence to statistical mechanics gives us several tools.  Here we summarize the evidence that these multiple tools lead to consistent answers, so that we can be confident in our estimates.

Our first try at entropy estimation is based on the  heat capacity integration in Eq.~(\ref{sint}).   To begin, with $N=10,\, 20$ neurons, we {\em can} enumerate all $2^N$ states of the network and hence we can find the maximum entropy distributions exactly (with no Monte Carlo sampling).  From these distributions we can also compute the entropy exactly, and it agrees with the results of the heat capacity integration.  Indeed, there is good agreement for the entire distribution, with Jensen-Shannon divergence between exact maximum entropy solutions and solutions using our reconstruction procedure at $\sim 10^{-6}$.   As a second check, now useable for all $N$, we note that the entropy is zero at $T=0$, but $S=N\,{\rm bits}$ at $T=\infty$.  Thus we can do the heat capacity integration from $T=1$ to $T=\infty$ instead of $T=0$ to $T=1$, and we get essentially the same result for the entropy (mean relative difference of $8.8\E{-3}$ across 30 networks at $N=100$ and $N=120$).

Leaning further on the mapping to statistical physics, we realize that the heat capacity is a summary statistic for the density of states.  There are Monte Carlo sampling methods, due to Wang and Landau \cite{wanglandau} (WL), that aim specifically at estimating this density, and those allow us to compute the entropy from a single simulation run.  The results, in Fig~\ref{sf3}A, are in excellent agreement with the heat capacity integration.

K-pairwise models have the attractive feature that, by construction, they  match exactly the probability of the all-silent pattern, $P(K=0)$, seen in the data. As explained in the main text, this means that we can ``measure'' the partition function, $Z$, of our model directly from the probability of silence.  Then we can compute the average energy $\langle E\rangle$ from a single MC sampling run, and find the entropy for each network.   As shown in Figs~\ref{sf3}B and C, the results agree both with the heat capacity integration and with the Wang--Landau method, to an accuracy of better than $1\%$.

\begin{figure}[b]
\includegraphics[width=\linewidth]{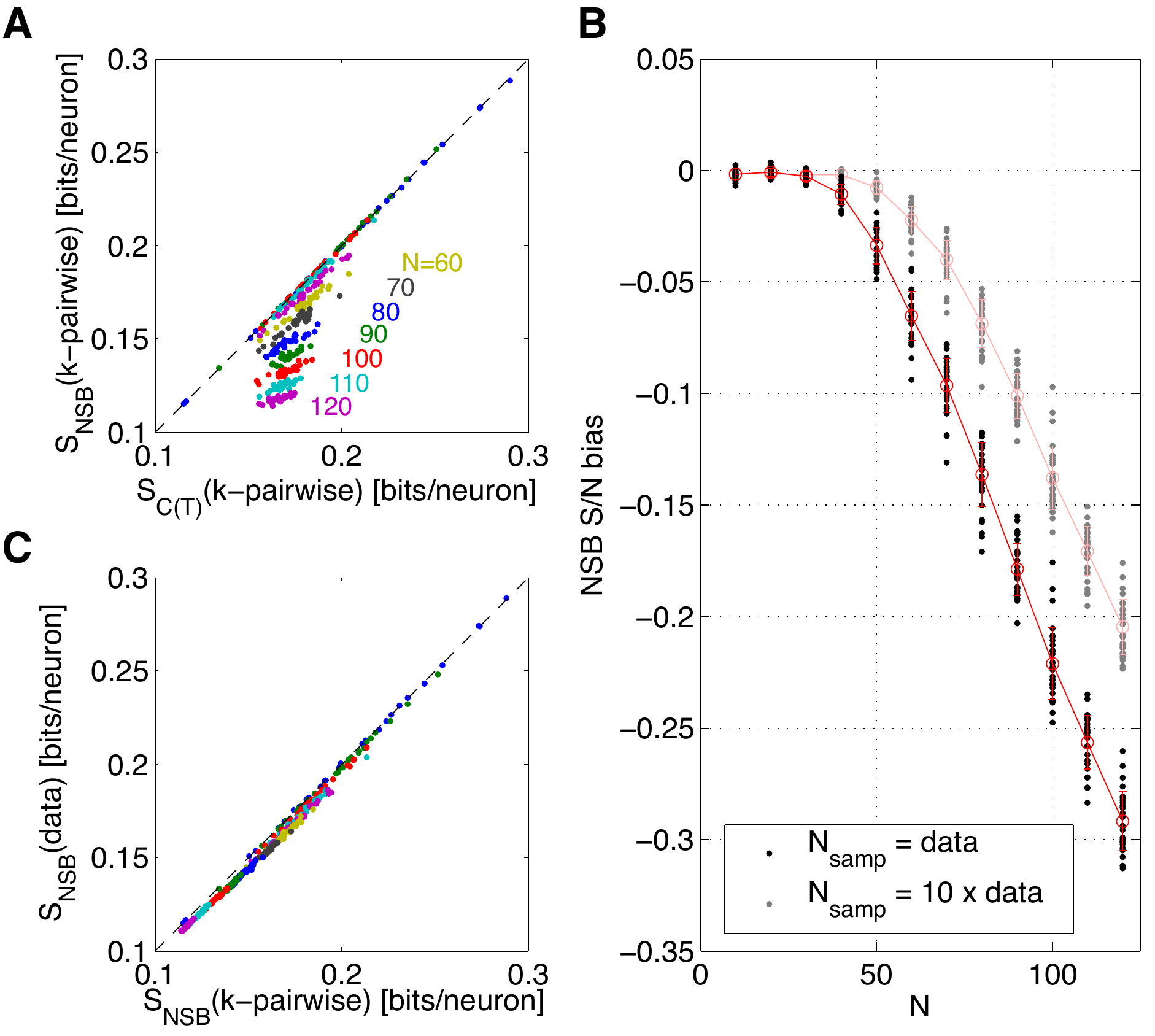} 
\caption{{\bf Sample-based entropy estimation.}{\bf (A)} The bias in entropy estimates computed directly from samples drawn from K-pairwise models. The NSB entropy estimate \cite{nsb} in bits per neuron computed using $\sim 3\E{5}$ samples from the model (same size as the experimental data set) on y-axis; the true entropy (using heat capacity integration) method on x-axis. Each dot represents one subnetwork of a particular size ($N$, different colors). For small networks ($N\leq 40$) the bias is negligible, but estimation from samples significantly underestimates the entropy for larger networks. {\bf (B)} The fractional bias of the estimator as a function of $N$ (black dots = data from (A), gray dots = using 10 fold more samples). Red line shows the mean $\pm$ s.d. over 30 subnetworks at each size. {\bf (C)} The NSB estimation of entropy from samples drawn from the model (x-axis) vs the samples from real experiment (y-axis); each dot is a subnetwork of a given size (color as in (A)). The data entropy estimate is slightly smaller than that of the model, as is expected for true entropy; for estimates from finite data this would only be expected if the biases on data vs MC samples were the same.
}
\label{sf2}
\end{figure}

Finally, there are methods that allow us to estimate entropy by counting samples even in cases where the number of samples is much smaller than the number of states \cite{nsb} (NSB).  The NSB method is not guaranteed to work in all cases, but the comparison with the entropy estimates from heat capacity integration (Fig \ref{sf2}A) suggests that so long as $N<50$, NSB estimates are reliable (see also \cite{berry+rava}). Figure~\ref{sf2}B shows that the NSB estimate of the entropy does not depend on the sample size for $N<50$; if we draw from our models a number of samples equal to the number found in the data, and then ten times more, we see that the estimated entropy changes by just a few percent, within the error bars.  This is another signature of the accuracy of the NSB estimator for $N<50$. As $N$ increases, these direct estimates of entropy become significantly dependent on the sample size, and start to disagree with the heat capacity integration.  The magnitude of these systematic errors depends on the structure of the underlying distribution, and it is thus interesting that NSB estimates of the entropy from our model and from the real data agree with one another up to $N=120$, as shown in Fig \ref{sf2}C.

\section{Are real networks in the perturbative regime?}
\label{App:perturb}

The pairwise correlations between neurons in this system are quite weak.  Thus, if we make a model for the activity of just two neurons, treating them as independent is a very good approximation.  It might seem that this statement is invariant to the number of neurons that we consider---either correlations are weak, or they are strong.  But this misses the fact that weak but widespread correlations can have a non--perturbative effect on the structure of the probability distribution.  Nonetheless, it has been suggested that maximum entropy methods are successful only because correlations are weak, and hence that we can't really capture non--trivial collective behaviors with this approach \cite{roudi2}.

\begin{figure}[]
\includegraphics[width=\linewidth]{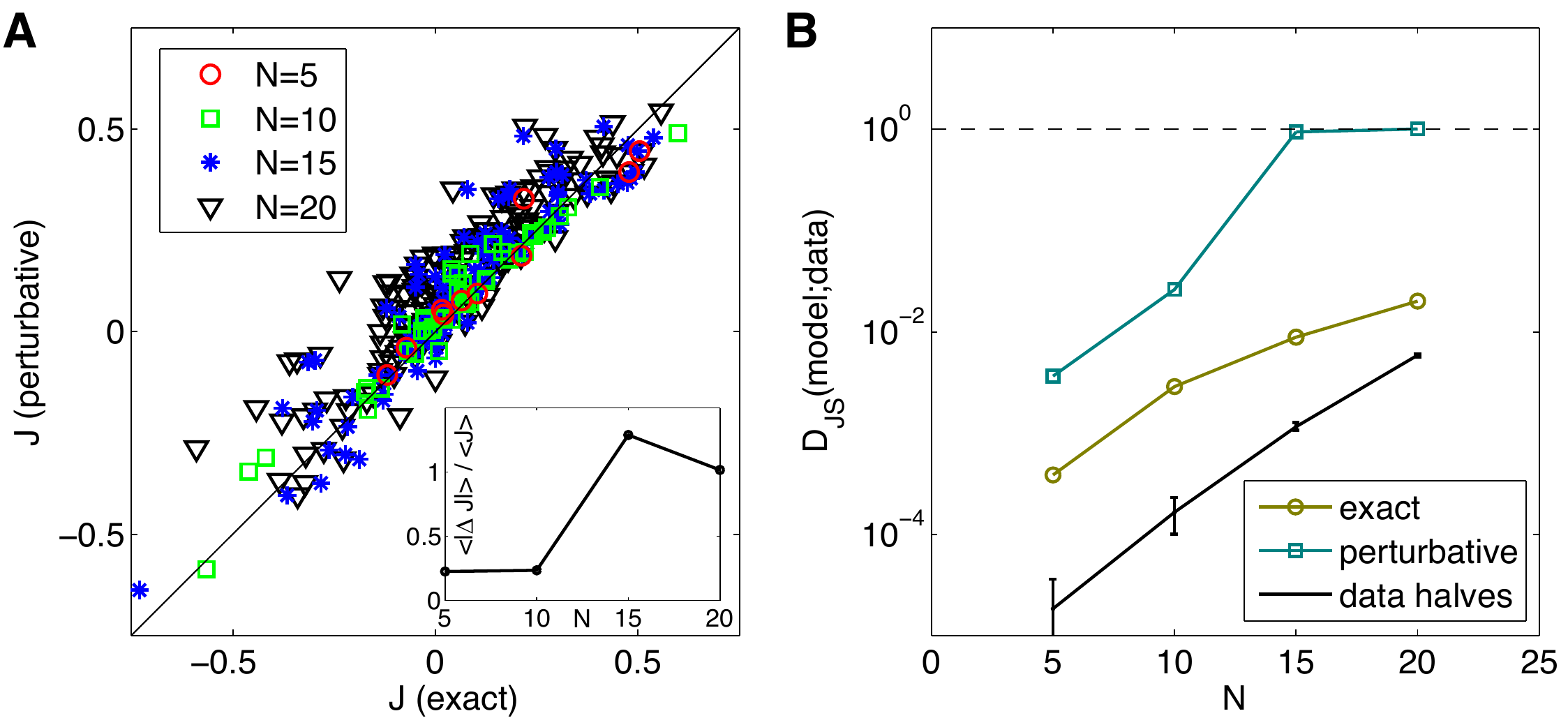} 
\caption{{\bf Perturbative vs exact solution for the pairwise maximum entropy models.} {\bf (A)} The comparison of couplings $J_{\rm ij}$ for a group of $N=5,10,15,20$ neurons,  computed using the exact maximum entropy reconstruction algorithm, with the lowest order perturbation theory result, $J_{\rm ij}=\frac{1}{4}\log c_{\rm ij}$, where $c_{\rm ij}=\langle\tilde{\sigma}_{\rm i}\tilde{\sigma}_{\rm j}\rangle/(\langle\tilde{\sigma}_{\rm i}\rangle\langle\tilde{\sigma}_{\rm j}\rangle)$ and $\tilde{\sigma}_{\rm i}=0.5(1+\sigma_{\rm i})$ \cite{roudi2,sessak}. In the case of larger networks, the perturbative $J_{\rm ij}$ deviate more and more from equality (black line). Inset: the average absolute difference between the true and perturbative coupling, normalized by the average true coupling. {\bf (B)} The exact pairwise model, Eq~(\ref{Hpair}), can be compared to the  distribution $P_{\rm expt}(\left\{\sigma_{\rm i}\right\})$, sampled from data; the olive line shows the Jensen-Shannon divergence (corrected for finite sample size) between the two distributions, for four example networks of size $N=5,10,15,20$. The blue line shows the same comparison in which the pairwise model parameters, $\vek{g}=\left\{h_{\rm i},J_{\rm ij}\right\}$, were calculated perturbatively. The black line shows the $D_{JS}$ between two halves of the data for the four selected networks.
\label{sf4}}
\end{figure}

While  independent models fail to explain the behavior of even small groups of neurons \cite{schneidman+al_06}, it is possible that groups of neurons might be in a weak perturbative regime, where the contribution of pairwise interactions could be treated as a small perturbation to the independent Hamiltonian, if the expansion was carried out in the correct representation \cite{roudi2}. 
 Of course, with finite $N$, all quantities must be analytic functions of the coupling constants, and so we expect that, if carried to sufficiently high order, any perturbative scheme will converge---although this convergence may become much slower at larger $N$, signaling genuinely collective behavior in large networks.

To make the question of whether correlations are weak or strong precise, we ask whether we can approximate the maximum entropy distribution with the leading orders of perturbation theory.  There are a number of reasons to think that this won't work \cite{monasson,monasson2,azhar_thesis,azhar}, but in light of the suggestion from Ref \cite{roudi2} we wanted to explore this explicitly.  If correlations are weak, there is a simple relationship between the correlations $C_{\rm ij}$ and the corresponding interactions $J_{\rm ij}$ \cite{roudi2, sessak}.  We see in Fig \ref{sf4}A that this relationship is violated, and the consequence is that models built by assuming this perturbative relationship are easily distinguishable from the data even at $N=15$ (Fig \ref{sf4}B).  We conclude that treating correlations as a small perturbation is inconsistent with the data.  Indeed, if we try to compute the entropy itself, it can be shown that even going out to fourth order in perturbation theory is not enough once $N> 10$ \cite{azhar_thesis,azhar}.

 \end{document}